\documentclass{aa}

\usepackage{graphicx}
\usepackage{txfonts}
\usepackage{amsmath,amssymb}
\usepackage{mathrsfs}   
\usepackage{mathtools}  
\usepackage{color}
\usepackage[bookmarks=true]{hyperref}
\hypersetup{
  colorlinks=true,
  pdfborder=2 2 0.,
  linkcolor=red,
  anchorcolor=black,
  citecolor=blue,
  urlcolor=black,
}
\usepackage{breakurl}

\usepackage{natbib}
\bibpunct{(}{)}{;}{a}{}{,} 

\newcommand{\Msun}{M_\odot}
\newcommand{\de}{{\rm d}}
\newcommand{\bx}{{\bf x}}
\newcommand{\bv}{{\bf v}}
\newcommand{\bJ}{{\bf J}}
\newcommand{\bL}{{\bf L}}
\newcommand{\lk}{\mathcal{L}}
\newcommand{\Jr}{J_R}
\newcommand{\Jphi}{J_\phi}
\newcommand{\Jz}{J_z}
\newcommand{\vr}{v_R}
\newcommand{\vphi}{v_\phi}
\newcommand{\vz}{v_z}

\newcommand{\fcomp}{f_{\rm comp}}
\newcommand{\fgal}{f_{\rm galaxy}}
\newcommand{\fthin}{f_{\rm thin}}
\newcommand{\fthick}{f_{\rm thick}}
\newcommand{\fhalo}{f_{\rm halo}}

\newcommand{\mhalo}{m_{\rm halo}}

\newcommand{\Mhalo}{M_{\rm halo}}
\newcommand{\Mdm}{M_{\rm DM}}
\newcommand{\Mcomp}{M_{\rm comp}}
\newcommand{\Mgal}{M_{\rm gal}}

\newcommand{\btheta}{\boldsymbol{\theta}}
\newcommand{\Gaia}{\textit{Gaia}}
\newcommand{\Tycho}{\textit{Tycho}}
\newcommand{\ra}{{\rm RA}}
\newcommand{\dec}{{\rm DEC}}
\newcommand{\mura}{\mu_{\rm RA}}
\newcommand{\mudec}{\mu_{\rm DEC}}
\newcommand{\vlos}{v_{\rm los}}
\newcommand{\cov}{\mathsf{C}}

\newcommand{\Jsun}{J_\odot}

\begin{document}

   \title{The dynamically selected stellar halo of the Galaxy with
                  \Gaia~and the tilt of the velocity ellipsoid}

   \titlerunning{The stellar halo with \Gaia~and RAVE}

   \author{Lorenzo Posti\inst{1}\fnmsep\thanks{posti@astro.rug.nl},
          Amina Helmi\inst{1},
          Jovan Veljanoski\inst{1}
          \and
          Maarten A. Breddels\inst{1}
          }

   \institute{Kapteyn Astronomical Institute, University of Groningen,
                          P.O. Box 800, 9700 AV Groningen, the Netherlands
             }

   \date{Received XXX; accepted YYY}


  \abstract
   {}
   {We study the dynamical properties of halo stars located in the
     solar neighbourhood. Our goal is to explore how the properties of
     the halo depend on the selection criteria used to define a
     sample of halo stars. Once this is understood, we proceed to
     measure the shape and orientation of the halo's velocity
     ellipsoid and we use this information to put constraints on the
     gravitational potential of the Galaxy.}
   {We use the recently released \Gaia~DR1 catalogue cross-matched to
     the RAVE dataset for our analysis. We develop a dynamical
     criterion based on the distribution function of stars in various
     Galactic components, using action integrals to identify halo members,
     and we compare this to the metallicity and to kinematically selected samples.}
   {With this new method, we find 1156 stars in the solar neighbourhood that are likely members
     of the stellar halo. Our dynamically selected sample consists
     mainly of distant giants on elongated orbits.  Their metallicity
     distribution is rather broad, with roughly half of the stars
     having [M/H]~$\ge -1$~dex. The use of different selection criteria
     has an important impact on the characteristics of the velocity
     distributions obtained. Nonetheless, for our dynamically selected
     and for the metallicity selected samples, we find the local
     velocity ellipsoid to be aligned in spherical coordinates in a Galactocentric
     reference frame. This suggests that the total
     gravitational potential is rather spherical in the region spanned
     by the orbits of the halo stars in these samples.}
   {}

   \keywords{Galaxy: kinematics and dynamics -- Galaxy: structure -- 
                         Galaxy: halo -- solar neighbourhood}

   \maketitle
%

\section{Introduction}

The study of our Galaxy offers a unique opportunity to unravel how
galaxies in general came about. In the Milky Way, the oldest stars
that we see today  formed from very pristine material and likely
still retain the memory of the physical and dynamical conditions of the
interstellar medium where they formed more than 10 Gyr ago
\citep[e.g.][for a review of the subject] {Eggen+1962,Helmi2008}.
These ``fossil'' stars are part of the so-called stellar halo of the
Galaxy, however their number is just a small fraction of the
population of stars that are present in the Galactic disc(s). Also
near the Sun, their spatial density is negligible. The present paper
revolves on how to find these ``precious'' stars in the context of the
ongoing ``golden era'' of Galactic surveys, which is characterized by
the advent of the revolutionary astrometric mission \Gaia~and vast
spectroscopic surveys such as the RAdial Velocity Experiment
\citep[RAVE,][]{Steinmetz+2006}, the Apache Point Observatory Galactic
Evolution Experiment \citep[APOGEE,][]{Majewski+17}, and in the near
future WEAVE \citep{Dalton+12} and 4MOST \citep{deJong+12}.

Halo stars are also interesting from the dynamical point of view
because they have elongated orbits that probe the outer regions of the
Galaxy, and thus can be used to study the total gravitational
potential of the Milky Way, including its dark matter
distribution. One measurable property that is often used to constrain
the shape of the dark matter halo is the orientation of the stellar velocity
ellipsoid \citep[e.g.][]{LyndenBell62,Ollongren62}. The most recent estimates
of the so-called {tilt} of the stellar velocity ellipsoid of both the
Galactic disc \citep{Siebert+2008,Smith+2012,Budenbender+2015} and the
stellar halo \citep{Smith+2009,Bond+2010,Carollo+2010,Evans+16} find that the
stellar velocity ellipsoid is aligned in a Galacto-centric spherical reference
frame and that it is elongated towards the Galactic centre. 
Although there has been some debate on its actual constraining
power \citep[e.g.][]{BinneyMcMillan2011}, recently \cite{AnEvans2016}
have established that  the shape of the total gravitational
potential can  indeed be locally constrained.

The  quest for stars in the Galactic halo is crucially influenced by
the prior knowledge that we put in the criteria to select these
objects. For instance, if we want to find the oldest stars in the
Galaxy, we might select stars by their scarcity of metals and we may
find that some of these stars move on high angular momentum near
circular planar orbits that are more typical of the disc
\citep[e.g.][]{Norris+1985}. On the other hand, if we want to find
stars with halo-like orbits, we might select the fastest moving stars
in a catalogue and encounter not only low-metallicity stars, but also more
metal-rich stars \citep[e.g.][]{Morrison+90,RyanNorris91}. Such
\emph{metallicity selected} or \emph{kinematically selected} samples
are biased by construction and this may affect the study of the
structural properties of the Galactic stellar halo. In general, it may
be preferable to identify stars using distribution functions. For
example, if the current position in the phase-space of a star is actually
known, then its full orbit can be reconstructed (given a Galactic
potential), and the impact of such biases can thus be limited by
analysing the stars that move far from the disc.

In this paper we  use the distribution function approach in order to
distinguish stars in the different Galactic components by their dynamics. We
 mainly use the fact that stars in the Galactic disc(s) are on
low-inclination, high angular momentum orbits,
while those in the stellar halo tend to have  significantly different
dynamics being on low angular momentum, eccentric, inclined
orbits. This will allow us to identify halo stars in the recently
released TGAS dataset \citep[which is part of the first \Gaia~data
release,][]{Gaia2016} in combination with the spectroscopic information
coming from the RAVE survey.  We will then use the resulting
metallicity  and kinematically unbiased sample of halo stars to
measure the tilt of the halo's velocity ellipsoid near the Sun. 

The paper is organized as follows. In Sect.~\ref{sec:data} we describe
the catalogue of stars considered from the TGAS and RAVE samples. In
Sect.~\ref{sec:method} we present the new method used to identify halo
stars given their dynamics. In Sect.~\ref{sec:results} we apply
these selection criteria to the stars common to TGAS and RAVE. We then
study the kinematics and metallicity distributions of the local
stellar halo and compare our results to those obtained using other
selection criteria.  We also measure in Sect.~\ref{sec:results} the
tilt of the velocity ellipsoid and discuss the implications on the
mass models of the Galaxy.  In Sect.~\ref{sec:concl} we present a summary
of our results and conclusions.

\section{Data: the TGAS-RAVE catalogue}
\label{sec:data}

The dataset used in this paper comes from the intersection of the TGAS
sample produced by the \Gaia~mission in its first data release
\citep{Gaia2016} and the stars observed by the RAVE Survey
\citep[DR5,][] {Kunder+2017}. We use the TGAS positions on the sky
and proper motions, together with their uncertainties and
mutual correlation. This sample has about two million stars with
magnitudes $6\lesssim G <13$. The radial velocities and
their uncertainties are from the RAVE DR5 catalogue, while
stellar parameters, such as surface gravities and metallicities,
and their uncertainties are from the updated RAVE pipeline by
\cite{McMillan+2017}. For the parallaxes $\varpi$, we consider
both the trigonometric values in the TGAS catalogue and the
spectro-photometric values derived by \cite{McMillan+2017}, and we use
for each star the one that has the smallest relative error
$\Delta\varpi/\varpi$, with $\Delta\varpi$ the parallax uncertainty.
The new determination by \cite{McMillan+2017} significantly improves 
the accuracy of the parallaxes of stars common to TGAS and RAVE
since it uses the trigonometric TGAS parallaxes as priors to
further constrain those derived spectro-photometrically with the
method of \cite{BurnettBinney2010}. 

From the RAVE DR5 catalogue we select only the stars satisfying the
following quality criteria:
\begin{itemize}
  \item[(i)] signal-to-noise of the RAVE spectrum S/N$\geq 20$;
  \item[(ii)] \cite{TonryDavis1979} correlation coefficient $>10$;
  \item[(iii)] flag for stellar parameter pipeline ${\verb+Algo_Conv_K+}\neq 1$;
  \item[(iv)] error on radial velocity $\epsilon_{RV} \leq 8$ km/s
\end{itemize}
\citep[see][]{Kordopatis+2013,Helmi+2017}. This ensures that the stellar
parameters, such as surface gravity, effective temperature, and metallicity,
and thus also the absolute magnitude, distance, and parallax, are well
constrained by the observed spectrum.
We use this subset of the RAVE stars to make our own cross-match with TGAS
based on the \Tycho-2 IDs, and find 185,955 stars in common. We further require
that
\begin{equation}\label{eq:cut_rpe}
\frac{\Delta\varpi_{\rm TGAS}}{\varpi_{\rm TGAS}} \leq 0.3 \quad {\rm or} \quad
  \frac{\Delta\varpi_{\rm RAVE}}{\varpi_{\rm RAVE}} \leq 0.3,
\end{equation}
where $\varpi_{\rm TGAS}$ is the measured TGAS parallax and $\varpi_{\rm RAVE}$
is the maximum likelihood value of the parallax, and $\Delta\varpi_{\rm TGAS}$ and $\Delta\varpi_{\rm RAVE}$ are their uncertainties. 
This gives 178,067 stars, of which  19,273 have 
$\Delta\varpi_{\rm TGAS}/\varpi_{\rm TGAS}<
\Delta\varpi_{\rm RAVE}/\varpi_{\rm RAVE}$, thus
we compute the distance to the star from the trigonometric TGAS parallax as 
$d=1/\varpi_{\rm TGAS}$; instead, for  the remaining 158,794 stars, 
we use the distance estimate from \cite{McMillan+2017}. Although 
 caution is required \citep[e.g.][]{ArenouLuri1999,BailerJones2015,
AstraatmadjaBailerJones2016}, recent analyses have shown that the
reciprocal of the parallax is actually the most accurate distance estimator
for stars with relative parallax error smaller than a few tens of percent 
\citep[][]{Binney+2014,SchonrichAumer2017}.
This is probably preferable than, for example,  the use of  a prior in the form of
an exponential for the intrinsic spatial distribution of halo stars 
(which are known to follow a power-law distribution).

More than $90\%$ of the halo stars we identify in our analyses below have
parallaxes derived as in \cite{McMillan+2017} since their relative parallax
error is smaller than that given by TGAS. Moreover, more than $90\%$ of the
halo stars have $\Delta\varpi/\varpi>0.1$, thus the cut at $30\%$ relative
parallax error (Eq.~\ref{eq:cut_rpe}) proves
to be good balance between number of halo stars identified and accuracy.
We also assess the robustness of the results in this paper against different
estimates of the spectro-photometric parallaxes by repeating the analysis
on a sample of stars with parallaxes from the RAVE DR5 public catalogue
\citep{Kunder+2017}; in Appendix \ref{sec:app} we show that all of our
main results are not significantly altered by this choice.

Finally, we exclude stars with $d<0.1$ kpc to reduce contamination
from the local stellar disc. Our final catalogue is thus composed of
175,006 stars.

\section{Identification of a halo sample}
\label{sec:method}

\subsection{Summary of the method}
\label{sec:method_summary}

The  main idea of our identification method is to distinguish Galactic components
(thin disc, thick disc, and stellar halo) on the basis of the orbits of
the stars. For instance, we consider  a star on a very elongated or
highly inclined orbit is  likely to be from the stellar halo, whereas a star
on an almost circular planar orbit is likely to be part of the thin disc. We
do not use any criterion based on metallicities or colours, or directly
dependent on the velocities of the stars. 

We characterize the orbits
of stars by computing a complete set of three integrals of motion,
which we choose to be the actions $\bJ$ in an axisymmetric
potential. Each dynamical model that we use for each Galactic component is
specified by the distribution function (hereafter DF) $f$, which is a function
that gives the probability of finding a star at a given point in phase space
\citep[see e.g.][]{BT08}.
We formalize this via the following procedure:
\begin{itemize}
\item[(i)] We define a DF $\fcomp$ for each  Galactic component. These are functions
      of the actions  $\bJ$ of a star (hence its orbit); they add up to the total DF of
      the Galaxy
      \begin{equation}
      \label{eq:f_galaxy}
      \fgal(\bJ) = \sum_{\rm comp} \fcomp(\bJ)
      \end{equation}
      and we use them as error-free likelihoods for the membership estimation of each
      component;
\item[(ii)] We compute the self-consistent total gravitational potential $\Phi$ solving
      Poisson's equation for the total density \citep[see e.g.][]{Binney2014}
      $\rho(\bx) = \int \de\bv\,\fgal(\bJ)$;
\item[(iii)] we characterize the orbits of stars by defining a canonical transformation from
      position-velocity to action-angle coordinates, which depends on the total
      gravitational potential,
      \begin{equation}
      \label{eq:xv_to_thetaJ}
      (\bx,\bv) \xmapsto{\quad\Phi\quad} (\btheta, \bJ),
      \end{equation}
      and we compute it with the `St\"{a}ckel Fudge' following \cite{Binney2012a};
\item[(iv)] for each star, we sample the error distribution in the space of observables
      assuming it is a six-dimensional multivariate normal with mean and covariance as
      observed. We apply coordinate transformations to these samples, from observed to
      $(\bx,\bv)$ and then to $(\btheta, \bJ)$ (Eq. \ref{eq:xv_to_thetaJ}), in order to estimate
      the error distribution in action space $\gamma(\bJ)$;
\item[(v)] we define the likelihood $\lk_i$ that a star $i$  belongs to a given component $\eta$
      by convolving the resulting error distribution in action space with the DF
      of that component as in (i), hence
      \begin{equation}
      \label{eq:star_likelihood}
      \lk_i(\eta\,|\bJ) = (f_\eta\ast\gamma_i)\,(\bJ).
      \end{equation}
\end{itemize}

\subsection{Actions and  distribution function of the Galaxy}
\label{sec:df}

Actions $\bJ$ are the most convenient set of integrals of motion in
galaxy dynamics, one of the reasons being that they are adiabatic
invariants.  In an axisymmetric potential, the first action $\Jr$
quantifies the extent of the radial excursion of the orbit, the second
$\Jphi$ is simply the component of the angular momentum $\bL$ along
the $z$-direction (which is perpendicular to the symmetry plane),
while the third action $\Jz$ quantifies the excursions in that direction.

In general actions are not easy to compute except for separable
gravitational potentials. Nonetheless, many methods have been
developed in recent years for general axisymmetric and triaxial
potentials and they all yield consistent results for the cases of
interest in galactic dynamics \citep[for a recent review of the
methods see][] {SandersBinney2016}.  Throughout the paper we 
adopt the algorithm devised by \cite{Binney2012a} to compute
$\bJ=\bJ(\bx,\bv)$ for all the stars in our sample, given a Galactic
potential $\Phi$.  In short, the so-called St\"{a}ckel Fudge algorithm
computes the actions in a separable St\"{a}ckel potential which is a
\emph{local} approximation to the Galactic potential $\Phi$ in the
region explored by the orbit. While there is, in principle, no guarantee
that a global set of action-angle variables exists for the given
potential $\Phi$, such a local transformation can always be found.

We then follow \citet[][see also \citeauthor{ColeBinney2017} \citeyear{ColeBinney2017}]
{Piffl+2014} and define the DF of the Galaxy to be an analytic function of the three action
integrals $\bJ=(\Jr,\Jphi,\Jz)$, as the superposition of three components: the thin disc,
thick disc, and stellar halo. We write the final DF as
\begin{equation}
\label{eq:df_galaxy}
\fgal(\bJ) = \fthin(\bJ)+\fthick(\bJ)+\fhalo(\bJ).
\end{equation}
The parameters of the \emph{quasi-isothermal} DFs of the discs are
described in \citet{Piffl+2014}, and have been constrained using the
RAVE dataset and for the purpose of this paper, are kept fixed. The mass of
each component is $\Mcomp=(2\pi)^3\int\de\bJ\,\fcomp(\bJ)$, and they
add up to $\Mgal = 4.2\times 10^{10}\Msun$.  Since we are interested
in stars in the solar neighbourhood, there is no need to model the
distribution in phase-space of the central regions of the Galaxy,
within $\lesssim 5$ kpc, thus  in the final model we  only include an
axisymmetric bulge as an external, fixed component in the total
potential. Modelling  the effect of the bar on the orbits of stars
in the solar neighbourhood as well goes beyond the scope of this paper.


For the stellar halo we assume the DF is a power-law function of the actions, 
\begin{equation}
\label{eq:df_halo}
\fhalo(\bJ)\equiv\mhalo\left[1+g(\bJ)/J_0\right]^{\beta_\ast},
\end{equation}
where $\mhalo$ is a constant such that $\Mhalo=5\times 10^{8}\Msun$, $\beta_\ast=-4$, and
$J_0=500\,{\rm kpc\,km/s}$, and where
\begin{equation}
\label{eq:gJ}
g(\bJ) \equiv \Jr + \delta_\phi |\Jphi| + \delta_z \Jz
\end{equation}
is a homogeneous function of the three actions. For $(\delta_\phi,\delta_z)=(1,1)$,
the model is nearly spherical in the solar neighbourhood. The constant $J_0$ serves
to produce a core in the innermost $R\lesssim 1.5$ kpc of the halo and it has the only
purpose of making the stellar halo mass finite.

This choice for the DF generates models with density distributions
which are also power laws \citep[][]{Posti+2015,WilliamsEvans2015};
for our choice of $\beta_\ast$ the density distribution of the halo
follows roughly $\rho\propto r^{-3.5}$.  This yields a reasonable
description for the stellar halo in the solar neighbourhood
(where our dataset is located), but does
not take into account the possibility that the structure of the
halo may change as a function of distance
\citep[e.g.][]{Carollo+2007,Deason+2011,DasBinney2016,Iorio+2017}.

Our  results are not very strongly dependent on the values chosen for
the characteristic parameters of the DF in Eq. \eqref{eq:df_halo}. We
have tested our method  $i$) with different geometries (spherical, as in
our default, or flattened, which for $(\delta_\phi,\delta_z)=(1/2,1)$
yields $q=0.7$ at $R_0$); $ii$) with different velocity distributions
(isotropic or tangentially/radially anisotropic); $iii$) with different
logarithmic slopes of the density distributions at $R_0$ (from -2.5 to
-3.5); and even $iv$) with different masses (from half to twice $\Mhalo$). We  found 
that more than $94\%$ of stars identified as `halo' are in
common for the different choices.

\subsection{Gravitational potential}
\label{sec:potential}

The Galactic potential that we use is made up of several massive
components of  two different kinds: i) static components and ii) 
DF components with self-gravity.

We have three components of the first kind, which we take from
\citet[][Table 1 and their results for the dark halo]{Piffl+2014}: a
gaseous disc, an oblate bulge, and an oblate dark matter halo.  The
gaseous disc is modelled as an exponential in both $R$ and $z$, with
scale-length $R_{\rm d, g}=5.36$ kpc and scale-height $z_{\rm d,g}=40$
pc, a $4$ kpc hole in the central region, and total mass of $M_{\rm
gas}=10^{10}\Msun$.  The bulge is instead an oblate ($q=0.5$) double
power-law density distribution with scale radius of $r_{\rm 0,b}=75$
pc, exponential cut-off at $r_{\rm cut,b}=2.1$ kpc, and total mass of
$M_{\rm bulge}=8.6\times 10^9\Msun$.  The dark matter is distributed
as a flattened ($q=0.8$) halo \citep{NFW96}, truncated at the virial
radius, with scale radius $r_{\rm 0,DM}=14.4$ kpc and virial mass of
$\Mdm=1.3\times 10^{12}\Msun$.

Each stellar component of the Galaxy that we modelled with a DF has a contribution
to the total gravitational potential $\Phi$ given by Poisson's equation. Starting
from an initial guess for the Galactic potential $\Phi_0$, which also contains  the
contributions from the static components, we compute the density distributions by
integrating their DFs over the velocities and then, using Poisson's equation, we compute
the new total galactic potential $\Phi_1$. We use the
iterative scheme suggested by \citet{Piffl+2015} to converge (in $\sim 5$ iterations)
to the total self-consistent gravitational potential $\Phi$.

With this procedure the final model, specified by this gravitational potential and
the galaxy DF given by Eq.~\eqref{eq:df_galaxy}, is self-gravitating, i.e. the
stars are not merely treated as tracers of the potential.

\subsection{Error propagation}
\label{sec:coords}

To compute the actions (and their uncertainties) from the observables,
we proceed as follows. We define a reference system centred on the
Galactic Centre, where the $z$-axis is aligned with the disc's angular
momentum, $x$-axis is aligned with the Sun's direction, and $y$-axis is positive
in the direction of rotation. In this frame, the Sun is located at
(-8.3, 0, 0.014)~kpc with respect to the Galactic Centre and we assume a
solar peculiar motion, with respect to the local standard of rest
(LSR), of $V_\odot=(11.1, 12.24, 7.25)$~km/s
\citep[][]{Schonrich+2010}.

In order to get an estimate of the error distribution in action space,
we start from the error distribution in the space of observables. We
assume that this can be described by a six-dimensional
multivariate normal distribution with the measurements as means, their
uncertainties as standard deviations, and their mutual correlations as
the normalized covariances.  For the sky coordinates and velocities
determined by TGAS $(\ra,\dec,\mura,\mudec)$, we have estimates of
their uncertainties and correlations, while the line-of-sight
velocities $\vlos$ only have uncertainty estimates from RAVE, i.e.
the correlations with the other coordinates are null. For those stars
for which we take the parallax $\varpi_{\rm TGAS}$ we include
correlations with the other TGAS coordinates in the covariance matrix,
while for those for which we use the $\varpi_{\rm RAVE}$ we set them
to be null.

We sample the resulting six-dimensional multivariate normal
distribution of each star with $5000$ discrete realizations, and
convert each realization to the desired reference frame where
we estimate the means, variances, and covariances. This completely
specifies the error distribution in the
new space if we further assume that  there the uncertainties are
also correlated Gaussians. We call $\gamma[{\bf q}, \cov({\bf q})]$ the
resulting multivariate distribution in the space of the new parameters
${\bf q}$ with covariance $\cov({}\bf q)$ and we refer to it 
as $\gamma({\bf q})$.

\subsection{Component membership estimation}
\label{sec:membership}

As discussed earlier, the orbit of a star in a given gravitational
potential is completely specified by the values of a complete set of
isolating integrals of motion such as the action integrals.  Suppose
now we have characterized the orbit of a star by computing the
actions, how do we evaluate the probability that a star belongs to one
of the Galactic components?

We answer this question by computing the phase-space probability
density of finding the given star with actions $\bJ$ for each of the three
stellar components, i.e. we compute the value of the DF. Since we are
interested in knowing which is the most likely component to which the
star belongs, we can work with relative quantities and we can define
likelihoods in terms of probability density ratios. In particular, we
proceed by defining for each component $\eta$ the likelihood
\begin{equation}
\label{eq:Pcomp_ef}
P_{\eta,\rm ef}(\bJ) \equiv \frac{f_\eta(\bJ)/M_\eta}
        {\displaystyle\sum_{\rm comp\ne\eta}\fcomp(\bJ)/\Mcomp},
\end{equation}
where the subscript `ef' indicates that it is an \emph{error-free} estimate.
If for the component $\eta$
\begin{equation}
\label{eq:memb_cond_ef}
P_{\eta,\rm ef}(\bJ)>1,
\end{equation}
then the probability of finding the star with actions $\bJ$ in that component is
greater than in the other components.

\begin{figure}
\includegraphics[width=.49\textwidth]{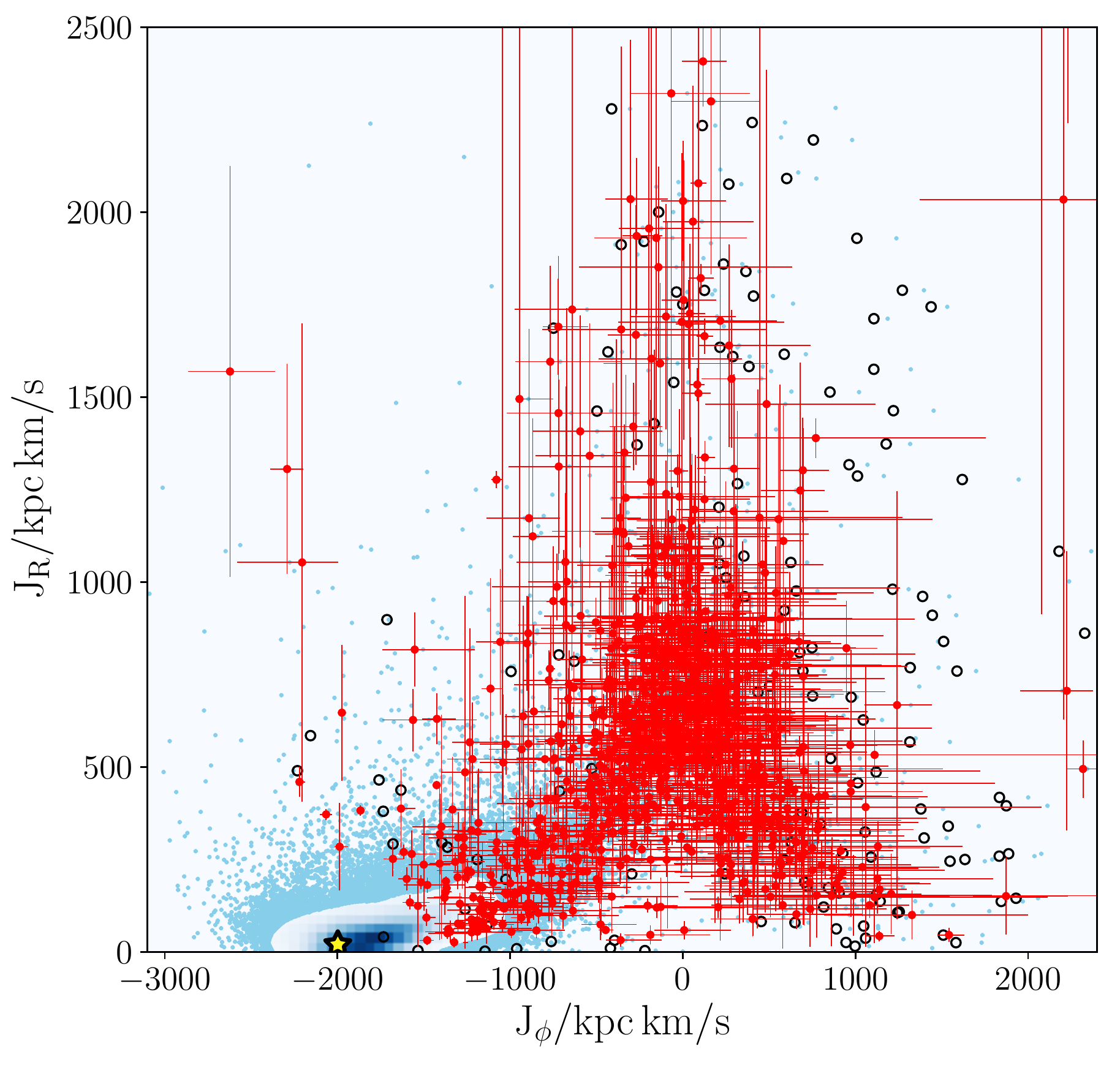}
\caption{Distribution of halo stars (red filled circles) in the
  angular momentum-radial action space. For comparison, we also plot
  all the stars in the sample (mostly disc stars) with cyan
  symbols and with a histogram of point density in the crowded region
  where orbits are nearly circular ($\Jr\sim 0$, $\Jphi\sim\Jsun$).
  The cyan points at large $\Jr$ and/or at small or retrograde
  $\Jphi$, although located in the region where the halo dominates,
   in practice have uncertainties that are too large and thus fail to pass the
  criterion given by Eq.~\eqref{eq:memb_cond}; they are therefore not
  part of the dynamically selected halo. The yellow star marks the
  position of the Sun in this diagram. We show 16th--84th percentile
  error bars, computed as in Sect. \ref{sec:coords}, for all the halo
  stars with fractional error on $\Jr$ smaller than $90\%$; instead,
  we plot as black empty circles the other halo stars for
  visualization purposes.}
\label{fig:Jr-Jphi_dynhalo}
\end{figure}

\begin{figure*}
\includegraphics[width=.33\textwidth]{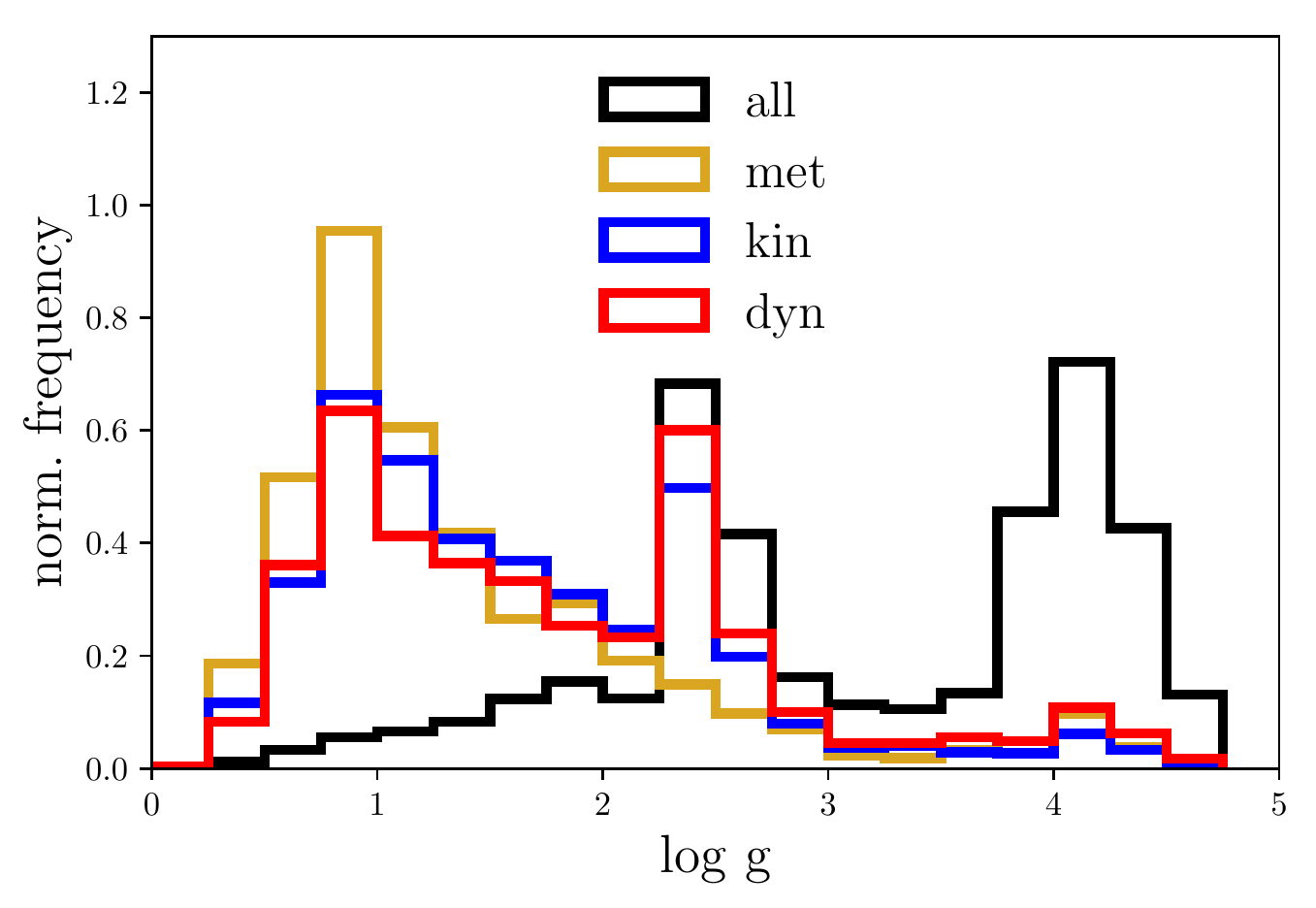}
\includegraphics[width=.33\textwidth]{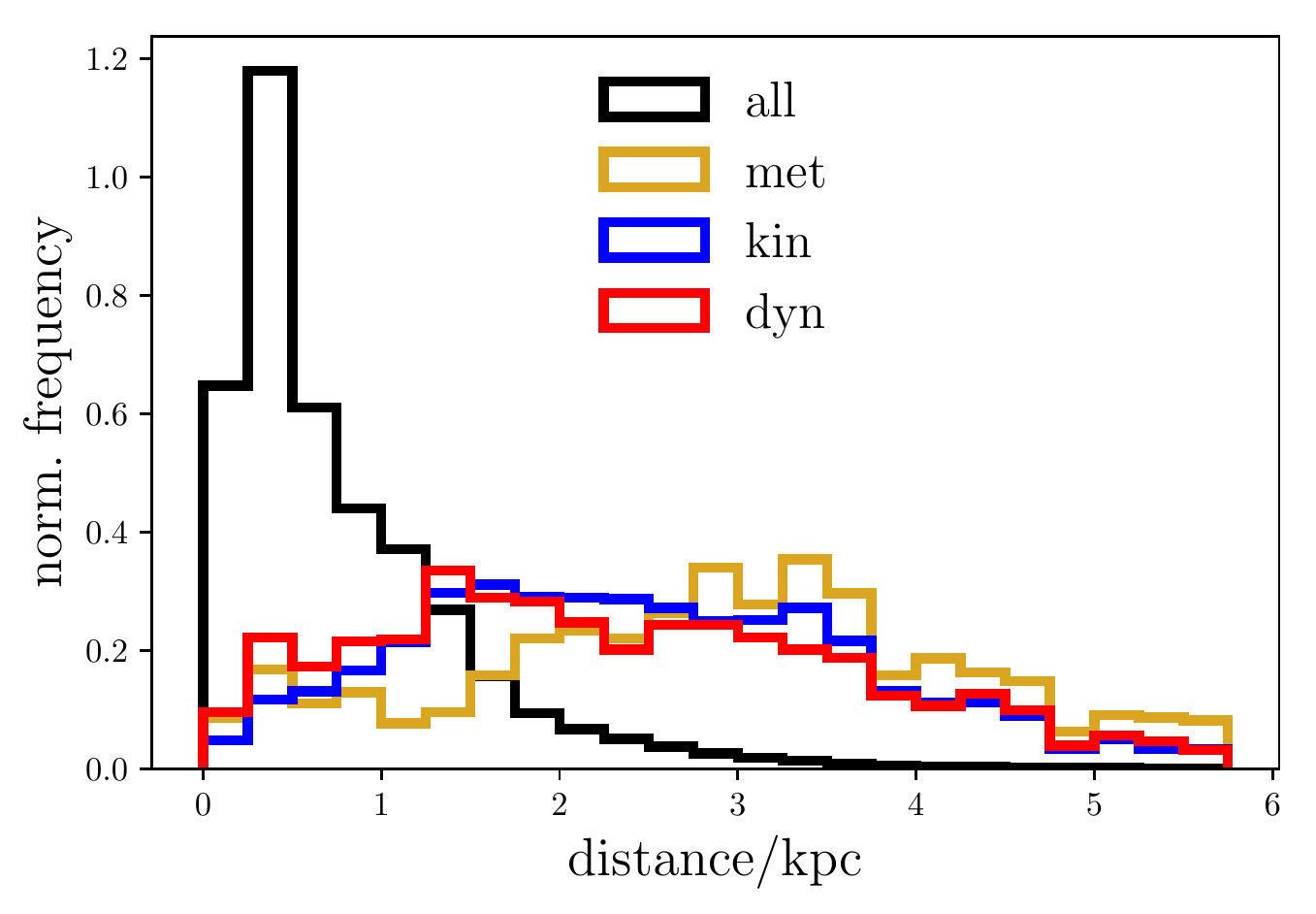}
\includegraphics[width=.33\textwidth]{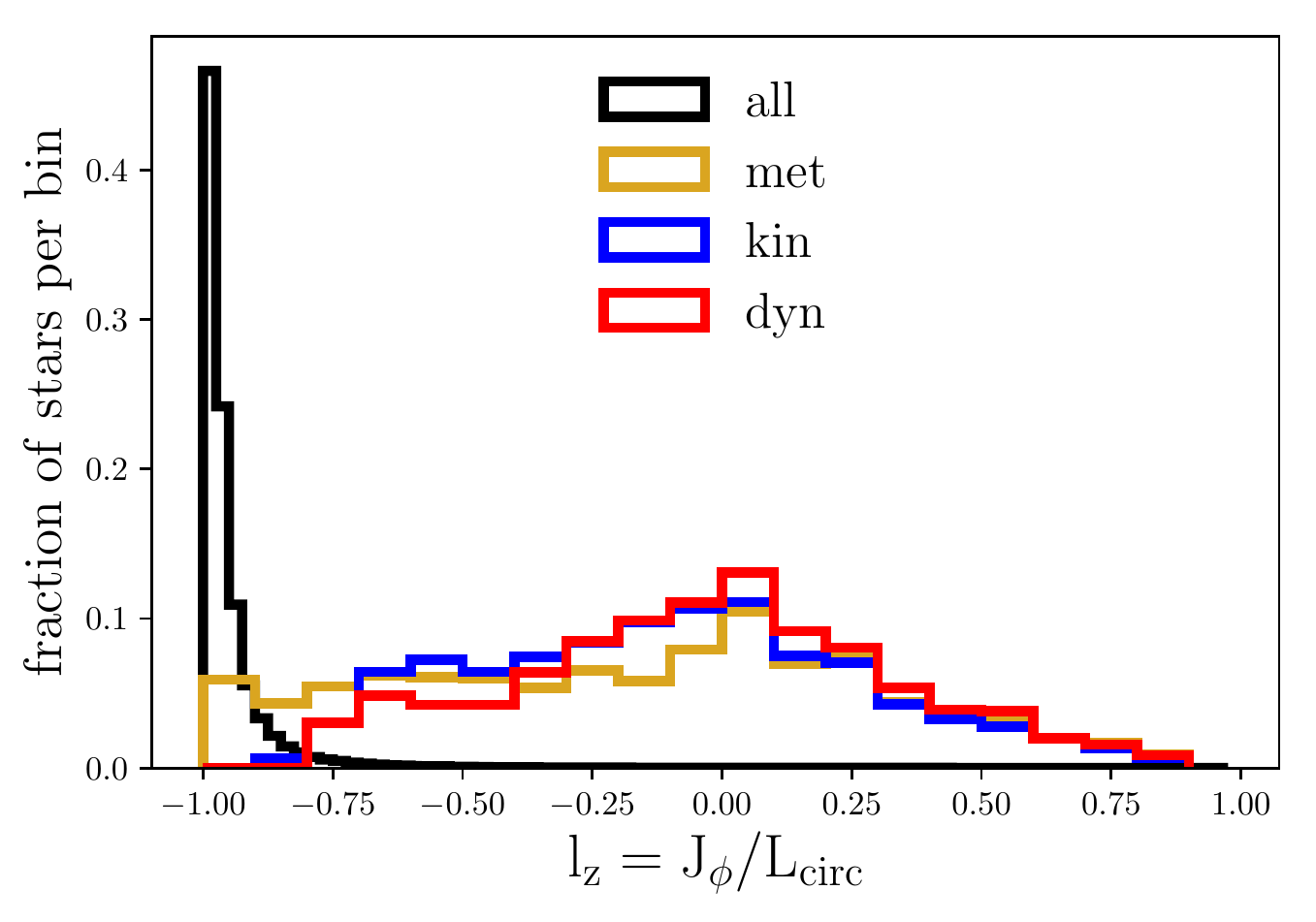}
\caption{Distribution of surface gravities (left), distances (middle), and circularities (right) of the stars
         in the dynamically selected (red),  kinematically selected (blue), and
         metallicity selected (yellow) stellar halo. For comparison, we also plot in black
         the distribution of all the stars in the TGASxRAVE sample defined in Sect. \ref{sec:data}.
         The circularities for the whole sample are shown on a 4x finer bin grid for
         visualization purposes.}
\label{fig:logg_dist_Distrib}
\end{figure*}

For each  star $i$, we estimate its error distribution in action space $\gamma_i(\bJ)$
starting from the error distribution of the observables as described in
Sect. \ref{sec:coords}.
We can take into account how measurement errors blur the estimate of the probability
density of each stellar component by convolving the DFs of each component $\eta$
with $\gamma_i$, hence
\begin{equation}
\label{eq:Pcomp}
P_\eta(\bJ) \equiv \frac{(f_\eta\ast\gamma_i)(\bJ)/M_\eta}
        {\displaystyle\sum_{\rm comp\ne\eta}(\fcomp\ast\gamma_i)(\bJ)/\Mcomp}.
\end{equation}
If this estimator is
\begin{equation}
\label{eq:memb_cond}
P_\eta(\bJ)>P_{\rm threshold}
\end{equation}
and $P_{\rm threshold} \ge 1$, then the probability that star $i$, with
actions $\bJ$ and error distribution in action space $\gamma_i$,
belongs to the component $\eta$ is higher than for any of the other
components.  The condition given by Eq.~\eqref{eq:memb_cond} is the
basis of our halo membership criteria.

\section{Results for the local stellar halo}
\label{sec:results}

We now apply the method described in the previous section to our
dataset. In order to have a prime sample of halo stars we used a more
conservative $P_\eta>P_{\rm threshold}$, and $P_{\rm threshold}=10$
proved to be a convenient choice.  Out of the $\sim 175000$ stars in
our dataset we find 1156  likely to belong to the stellar halo\footnote{
The spectra of the vast majority of halo stars (1104)
is classified as `normal' (${\it flag\_N}=0$) according to the spectroscopic
morphological classification by \cite{Matijevic+2012}. This ensures that
binaries and peculiar stars do not play a relevant role in our analysis
and that the radial velocities and the stellar parameters of halo stars
are safely determined. However, for the sake of making an honest comparison
with the halo sample selected by \cite{Helmi+2017}, we do not exclude
the 52 halo stars with peculiar spectra in what follows.
}
($0.66\%$). Of these, 784 are found within $d<3$ kpc.

In Figure \ref{fig:Jr-Jphi_dynhalo} we show the distribution of stars
identified as halo in a projection of the action-space, namely on
the $\Jphi-\Jr$ plane, and we compare it to the distribution of all
the other stars in the catalogue.  The vast majority of the stars in
the TGAS-RAVE catalogue are actually disc stars on nearly circular
orbits, hence they mostly have high angular momentum
($\Jphi\sim\Jsun\simeq-2000$ kpc km/s, where $\Jsun$ is the angular
momentum of the Sun), low  eccentricity ($\Jr\sim 0$), and typically
stay close to the plane ($\Jz\sim 0$).  The halo stars, instead, have
typically $\Jphi\sim 0$ and $\Jr\simeq 500-1000$ kpc km/s, meaning
that they are on low angular momentum and elongated orbits. For
completeness, in Fig.~\ref{fig:Jr-Jphi_dynhalo} we also show the
uncertainties on the action integrals that we estimate by computing the
interval of 16th--84th percentiles of the resulting discrete
samples obtained by propagating the measured error distribution in the
observables (Sect. \ref{sec:coords}).  The mean error on the
radial action for this sample is $\Delta\Jr\sim 125$ kpc km/s
($\Delta\Jr/\Jr\sim 26\%$), while that of the angular momentum is
$\Delta\Jphi\sim 280$ kpc km/s ($\Delta\Jphi/\Jphi\sim 17\%$) and that
of the vertical action is $\Delta\Jz\sim 80$ kpc km/s
($\Delta\Jz/\Jz\sim 37\%$).


\subsection{Comparison samples}

We  now analyse the local kinematics and the metallicity distribution
of the  dynamically selected halo. We  compare these distributions to
those of samples obtained by applying two widely used selection criteria
for halo stars: a selection based on metallicity
and one based on kinematics.

\subsubsection{ Metallicity selected local stellar halo}

A commonly used selection criterion for halo stars in the Galaxy is to
assume that they are typically old and formed of low-metallicity
material. The idea is then to identify as halo all the stars that are
more metal-poor than a given threshold, typically a small
percent  of the solar metallicity.  

For our comparison we use a sample of metallicity selected halo stars
compiled as in \cite{Helmi+2017}, but using the distances to
the stars from the updated pipeline by \citet[][Veljanoski et al. in
preparation]{McMillan+2017}.  The Galactic halo is defined as the
component  whose stars have metallicity $\rm[M/H]\leq-1$~dex
\citep[RAVE calibrated, see][]{Zwitter+2008}. Since this metallicity
threshold is not very strict, a second criterion is applied after a
two-Gaussian decomposition of the sample is done;  a star is said to
belong to the halo if its velocity is more likely compatible with a
normal distribution in $(V_X,V_Y,V_Z)$ centred at $V_Y\sim 20$ km/s
than with another centred at $V_Y\sim 180$ km/s \citep[for more
details, see Sect. 2.2 in][]{Helmi+2017}.  This sample has a total of
1217 stars, of which 713 are found closer than 3 kpc.

\subsubsection{Kinematically selected local stellar halo}

Halo stars are also commonly identified as the fastest moving stars with respect
to the LSR. In particular, following \cite{NissenSchuster2010},
we define the kinematically selected stellar halo sample as being comprised of all the stars
with $|{\bf V}-{\bf V}_{\rm LSR}|>{\bf V}_{\rm LSR}$, where ${\bf V}=(V_X,V_Y,V_Z)$ is
the star's velocity in Galactocentric Cartesian coordinates and ${\bf V}_{\rm LSR}=
(0,v_{\rm LSR},0)$ km/s is the velocity of the LSR \citep[see also][]{Bonaca2017}. For the Galactic potential we use
 $v_{\rm LSR}=232$ km/s (Sect.~\ref{sec:potential}). The sample comprises a total of
1956 stars, of which 1286 are found closer than 3 kpc.

\subsection{Properties of the different samples}

As   is clear from Figure~\ref{fig:logg_dist_Distrib}, where we plot
the distribution of stellar surface gravities $\log g$ (as measured by RAVE)
and distances $d$ for the three samples considered here, most stars
are distant giants. We do not see any significant bias or difference
in the $\log\,g$ and $d$ distributions (left and middle panels,
respectively) in the three samples. The right panel of
Fig.~\ref{fig:logg_dist_Distrib} shows the circularity $l_z$ for the
samples and also includes  the distribution for the whole TGAS-RAVE
sample.  We define the circularity $l_z$ of a star to be the ratio of
its angular momentum to that of the circular orbit at the same energy
$L_{\rm circ}(E)$, i.e. $l_z = \Jphi/L_{\rm
  circ}(E)$. Figure~\ref{fig:logg_dist_Distrib} shows that the proposed dynamical
selection criteria  successfully distinguishes halo stars from
those in the disc (for which $l_z \sim -1$). We also note  that the
metallicity selected sample contains stars with orbits that are
disc-like.

Very distant stars typically span different physical volumes than more closer ones.
The complex TGAS-RAVE selection function likely yields a rather incomplete sample
of stars at large distances, hence we will work with just a \emph{local} sample of
halo stars defined as those with $d\leq 3$~kpc for most of the following discussion.

\subsubsection{Local kinematics}
\label{sec:localkin}

\begin{figure*}
\begin{center}
\includegraphics[width=.33\textwidth]{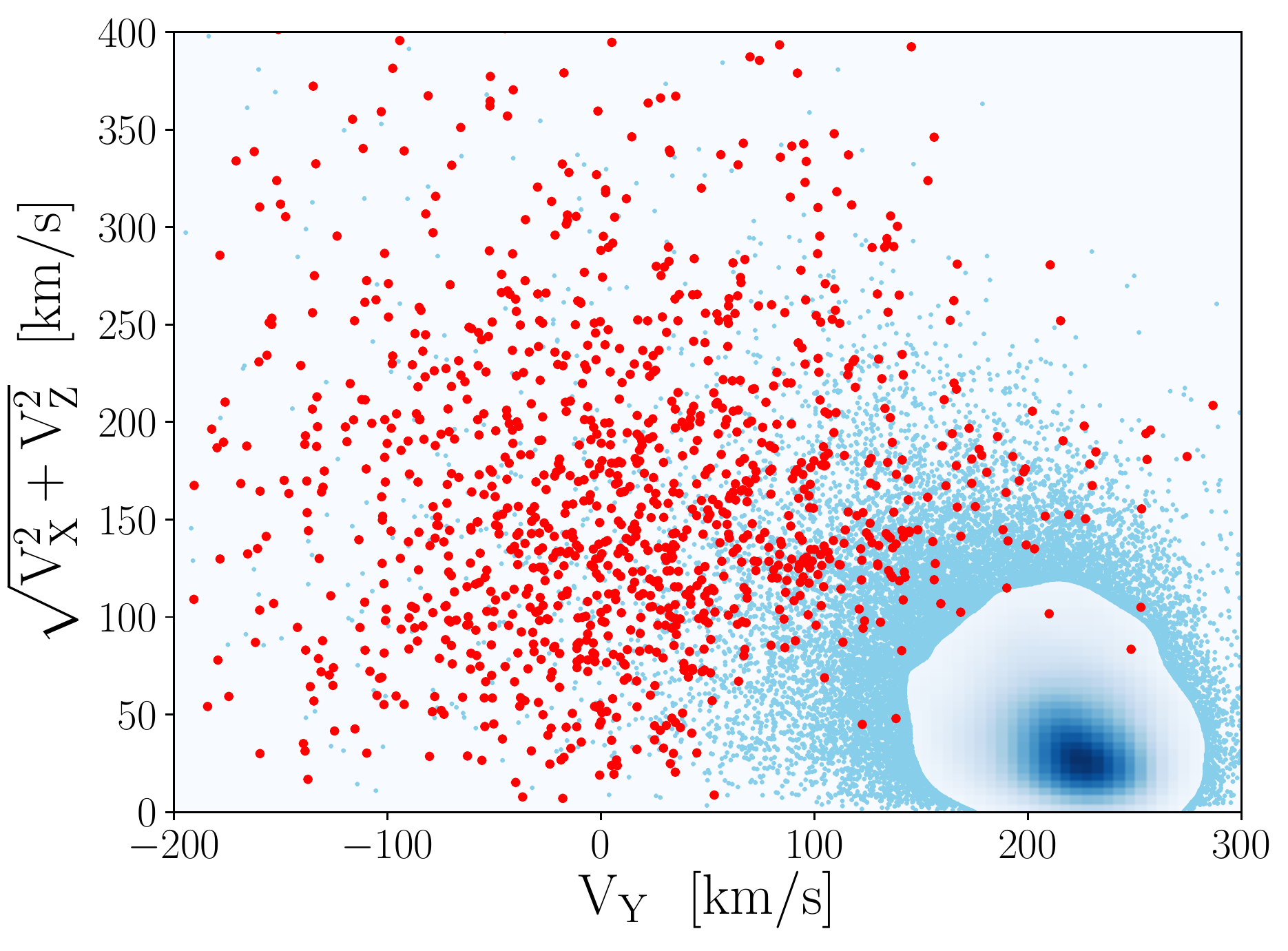}
\includegraphics[width=.33\textwidth]{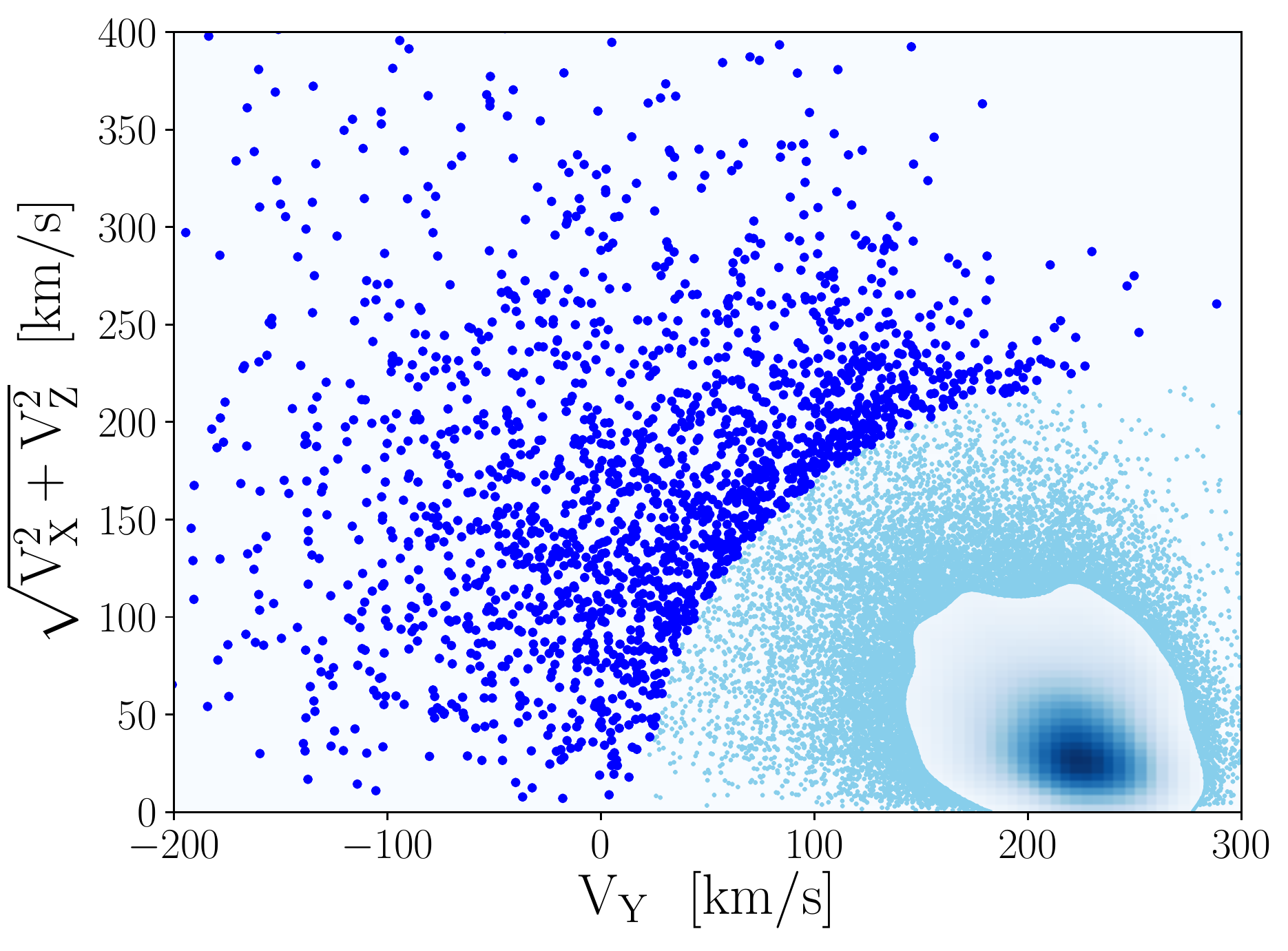}
\includegraphics[width=.33\textwidth]{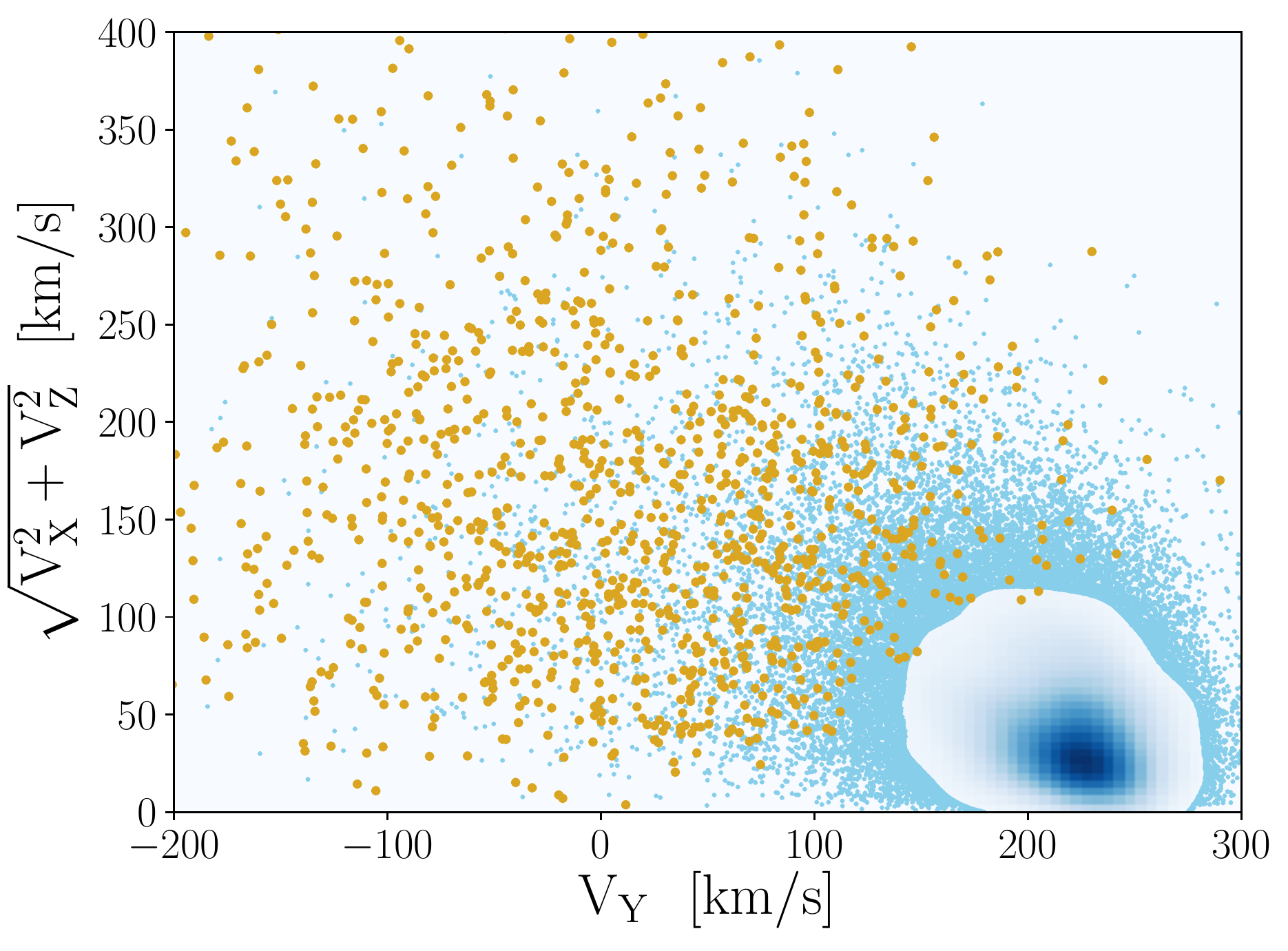}
\end{center}
\caption{Toomre diagram for the stars in our TGASxRAVE sample. In all the panels
  we plot all the stars with cyan symbols (and binned density). The
  red, blue, and yellow circles are respectively the stars in the
  dynamically selected, kinematically selected, and
  metallicity selected stellar halo.}
\label{fig:toomre}
\end{figure*}

In Galactocentric Cartesian coordinates $(V_X,V_Y,V_Z)$ we define the
 Toomre diagram as the plane $V_Y$ versus
$\sqrt{V_X^2+V_Z^2}$. This is shown in Figure~\ref{fig:toomre} for the
stars in the three selected halo samples.  The kinematically selected
halo is actually defined in this plane by the semi-circle $|{\bf
  V}-{\bf V}_{\rm LSR}|={\bf V}_{\rm LSR}$, which sets a sharp cut for
the halo/non-halo stars (Fig.~\ref{fig:toomre}, middle panel). In both
 the dynamically  and the metallicity selected samples there are, instead, a
non-negligible number of stars in the region that would not be allowed
by the kinematic criterion (respectively $18\%$ and $22\%$).  On the
other hand, of the halo stars selected by kinematics there are 1204
stars that are more metal-rich than [M/H]~$=-1$~dex, hence not belonging to the
metallicity selected sample, and 1009 stars that are not selected
dynamically because their measurement errors are too large to pass the
condition defined by Eq.~\eqref{eq:memb_cond} with our strict
threshold value. These two groups of stars are represented by the cyan
points outside the semi-circle $|{\bf V}-{\bf V}_{\rm LSR}|>{\bf
  V}_{\rm LSR}$ in the right and left panels of
Fig.~\ref{fig:toomre}, respectively.

\begin{figure*}
\includegraphics[width=\textwidth]{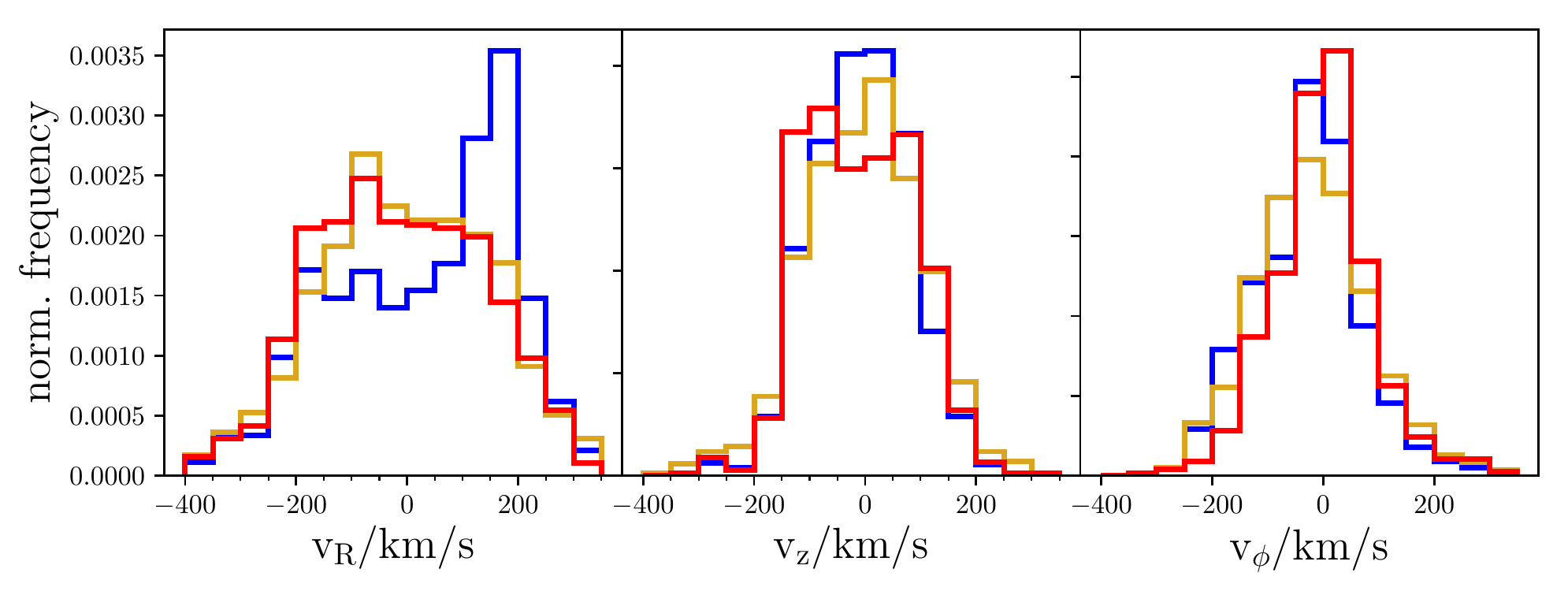}
\caption{One-dimensional distributions of the Galactocentric cylindrical velocities for stars in the dynamically selected (red), 
         kinematically selected (blue), and metallicity selected (yellow) stellar halo.}
\label{fig:velhist}
\end{figure*}
\begin{figure*}
\includegraphics[width=.95\textwidth]{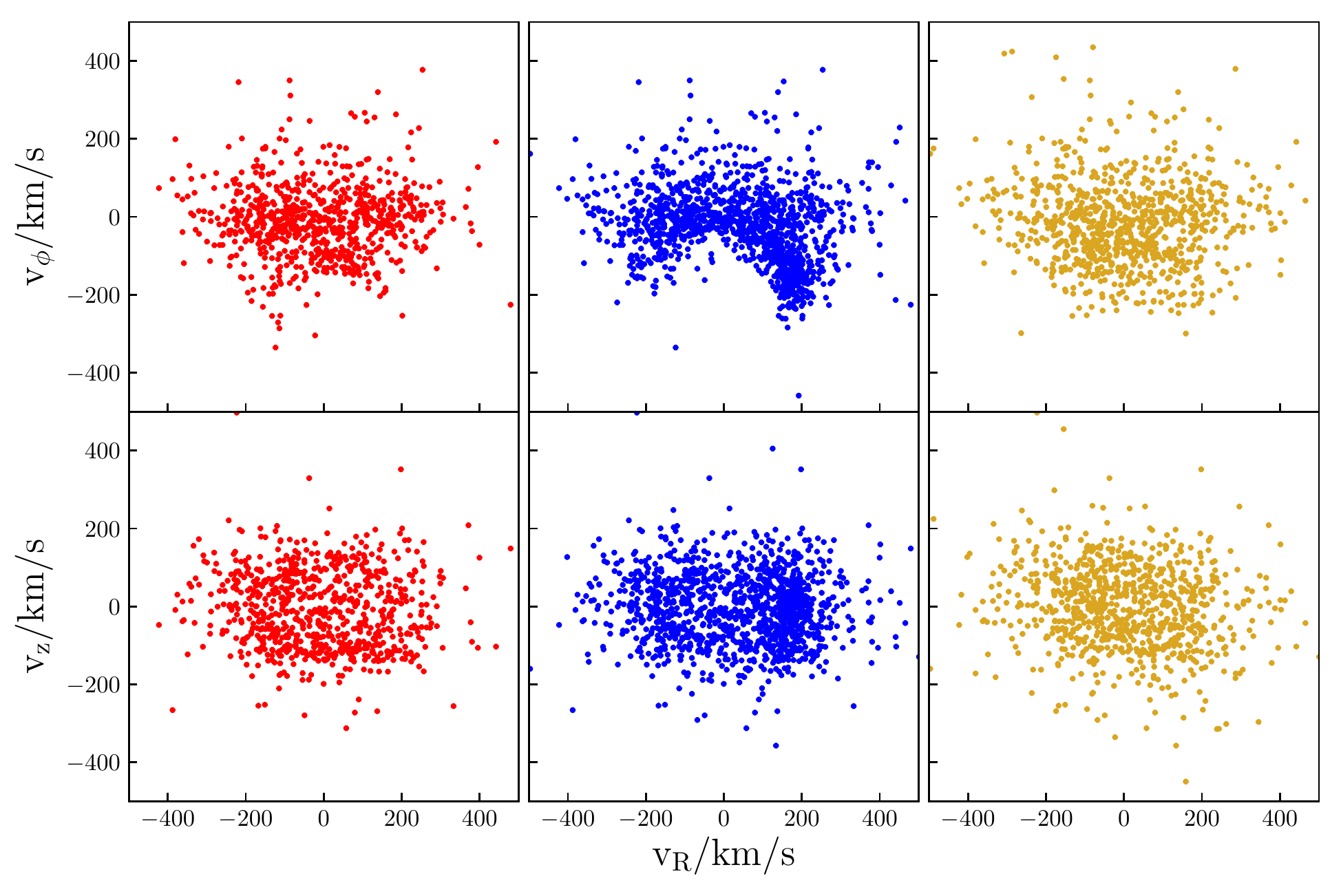}
\caption{Two-dimensional distributions of the Galactocentric cylindrical velocities
  for the dynamically selected (red),  kinematically selected
  (blue), and metallicity selected (yellow) stellar halo stars.}
\label{fig:localkin}
\end{figure*}

\begin{figure}
\includegraphics[width=.5\textwidth]{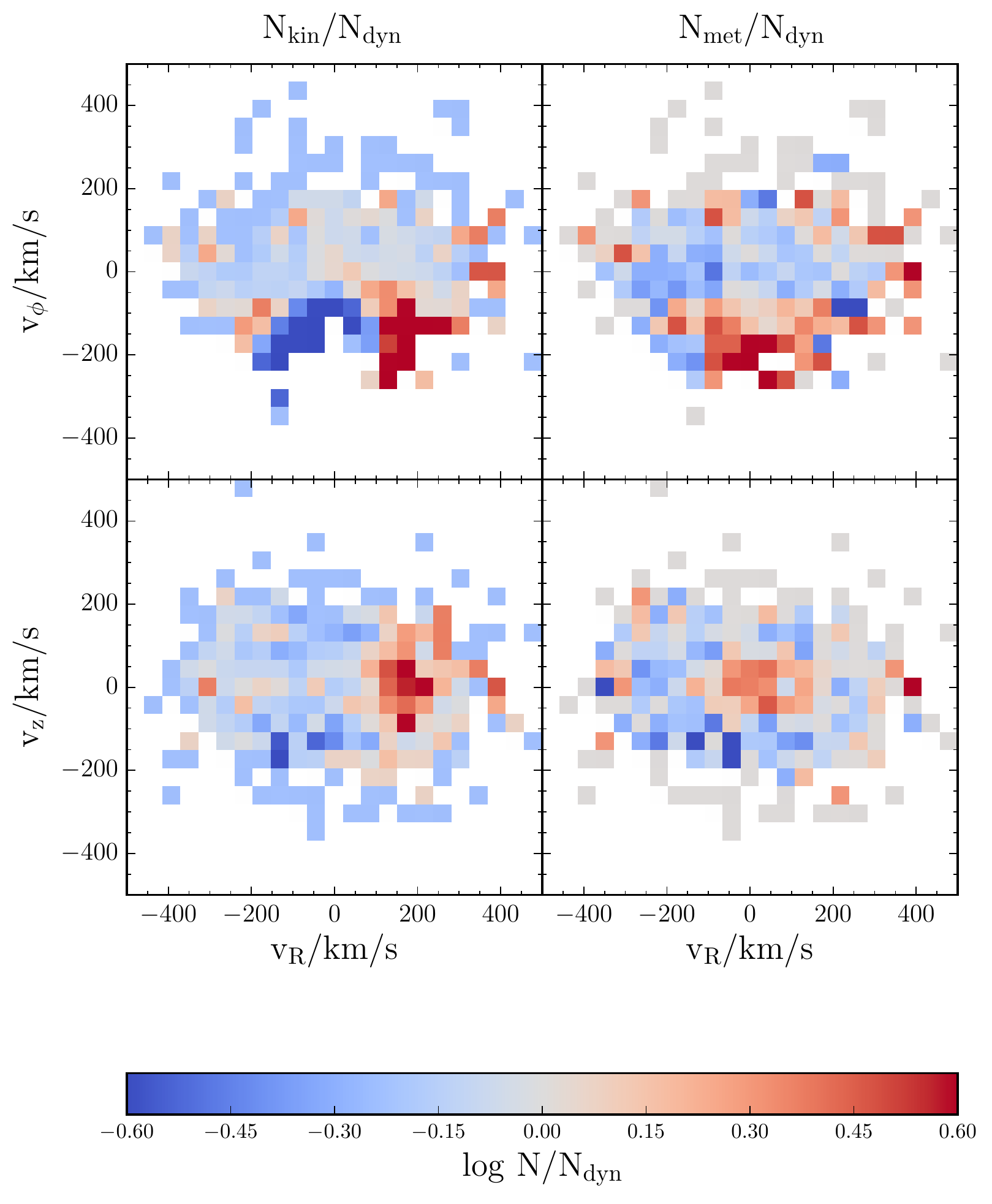}
\caption{Ratio of the normalized number of halo stars selected by kinematics
         ($\rm N_{kin}=N_{bin, kin}/N_{tot, kin}$, left panels) and by metallicity 
         ($\rm N_{met}=N_{bin, met}/N_{tot, met}$, right panels) to that of halo
         stars selected by dynamics ($\rm N_{dyn}=N_{bin, dyn}/N_{tot, dyn}$) in bins
         of the $\vr-\vphi$ (top) and $\vr-\vz$ (bottom) velocity subspaces. Bins are
         coloured by the logarithm of the number counts ratio.}
\label{fig:diff_vel}
\end{figure}


The kinematic selection of halo stars
produces unrealistic asymmetries in the plane of $V_Y$ and
$\sqrt{V_X^2+V_Z^2}$ (Fig.~\ref{fig:toomre}). These asymmetries translate into similar features
in the distribution of cylindrical Galactocentric velocities ($\vr$,
$\vphi$, $\vz$), as shown in Figure~\ref{fig:velhist}.  While the
histograms of the radial and azimuthal velocities of the
kinematically selected halo appear quite asymmetric, the distributions
for the dynamically selected halo are relatively symmetric, and are 
consistent with being Gaussian\footnote{Tested with   
d'Agostino's $K^2$ test.}.

In Figure~\ref{fig:localkin} we show the distributions of stars in the
$\vr-\vz$ and $\vr-\vphi$ planes for the three selected samples. 
To better highlight the differences in the velocity distributions in the
three cases, we bin velocity space uniformly and show in 
Figure~\ref{fig:diff_vel} the ratio of the fraction of stars per bin in the
kinematically  (or metallicity) selected halo to that of the dynamically selected
halo.
For the kinematically selected halo there is a significant lack of stars
close to $\vphi\sim-200$ km/s and $\vr \sim 0$, where most of the disc
stars are (blue region in the top left panel of
Fig.~\ref{fig:diff_vel}). This occurs because a kinematic selection is unable 
to distinguish between disc and halo stars in the region dominated by the
disc. Such a selection also produces a peak in the Galactocentric
radial velocity distribution at $\vr \simeq 180$ km/s (clearly visible
as a high-contrast red region in the top left and bottom left panels of 
Fig.~\ref{fig:diff_vel}). By exploring their dynamical properties we believe 
that these stars are likely thick disc contaminants since they have little 
vertical action, about $\Jz\simeq 35$ kpc km/s, which is close to the typical 
value of 10--20 kpc km/s for thick disc stars, and have typical circularity 
$l_z\simeq -0.7$, which is significantly closer to that of circular orbits
($l_z=-1$) in comparison to the rest of the halo sample.

Conversely, in the region of velocity space dominated by the disc
$(\vr,\vphi) \simeq(0,-200)$ km/s, there is a significant
concentration of stars in the case of a halo selected by metallicity
(red region in the top right and bottom right
panels of Fig.~\ref{fig:diff_vel}), indicating that there is still significant
residual contribution of the metal-poor tail of the thick disc. The distribution of circularities (right  panel of
Fig.~\ref{fig:logg_dist_Distrib}) also shows that there is indeed an excess of
stars with $l_z\lesssim -0.8$ with respect to the kinematically  and
dynamically selected samples.
These stars contribute to making the distribution of $\vphi$ for the
metallicity selected halo not centred on zero, but on a slightly prograde
value of $\overline{\vphi}=-25\pm 4$~km/s. If we remove  
stars with disc-like circularities (i.e.~$l_z\lesssim -0.8$) from this sample, 
then $\overline{\vphi}=-12\pm 6$ km/s \citep[see also][]
{Deason+2017,Kafle+2017}.

\subsubsection{Metallicity distribution}

\begin{figure}
\includegraphics[width=.49\textwidth]{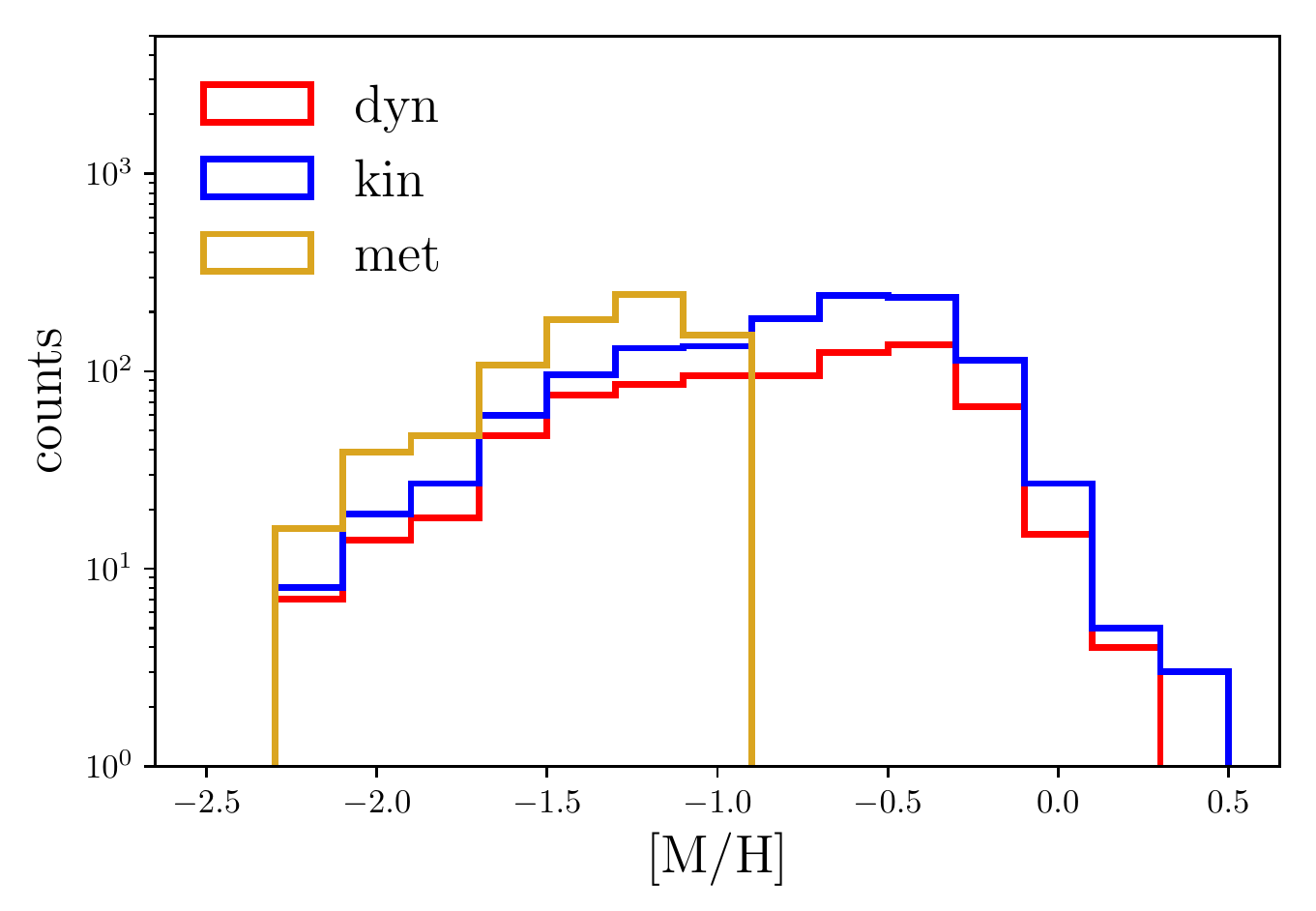}
\caption{Metallicity distribution (in number counts) for the dynamically selected (red), 
                 kinematically selected (blue), and metallicity selected (yellow) stellar halo samples.
         The RAVE calibrated metallicity [M/H] is from  \cite{McMillan+2017} and
         the typical uncertainty is $\sim 0.2$~dex.}
\label{fig:metDistrib}
\end{figure}

In Figure~\ref{fig:metDistrib} we show the RAVE calibrated metallicity distribution of
the halo stars for the three samples. While the metallicity selected halo has a sharp
edge at [M/H]~$\sim-1$~dex, which is the threshold for the selection, in the other
two cases the distribution is peaked at  [M/H]~$\sim-0.5$~dex and declines
smoothly at higher metallicities reaching [M/H]~$\sim 0$. In particular, we find
633 stars ($55\%$) with [M/H]~$>-1$~dex in the dynamically selected
sample. These stars are all on highly elongated, low angular momentum orbits, and
215 of them ($34\%$) are on retrograde orbits \citep[see also][]{Bonaca2017}.

\begin{figure}
\includegraphics[width=.5\textwidth]{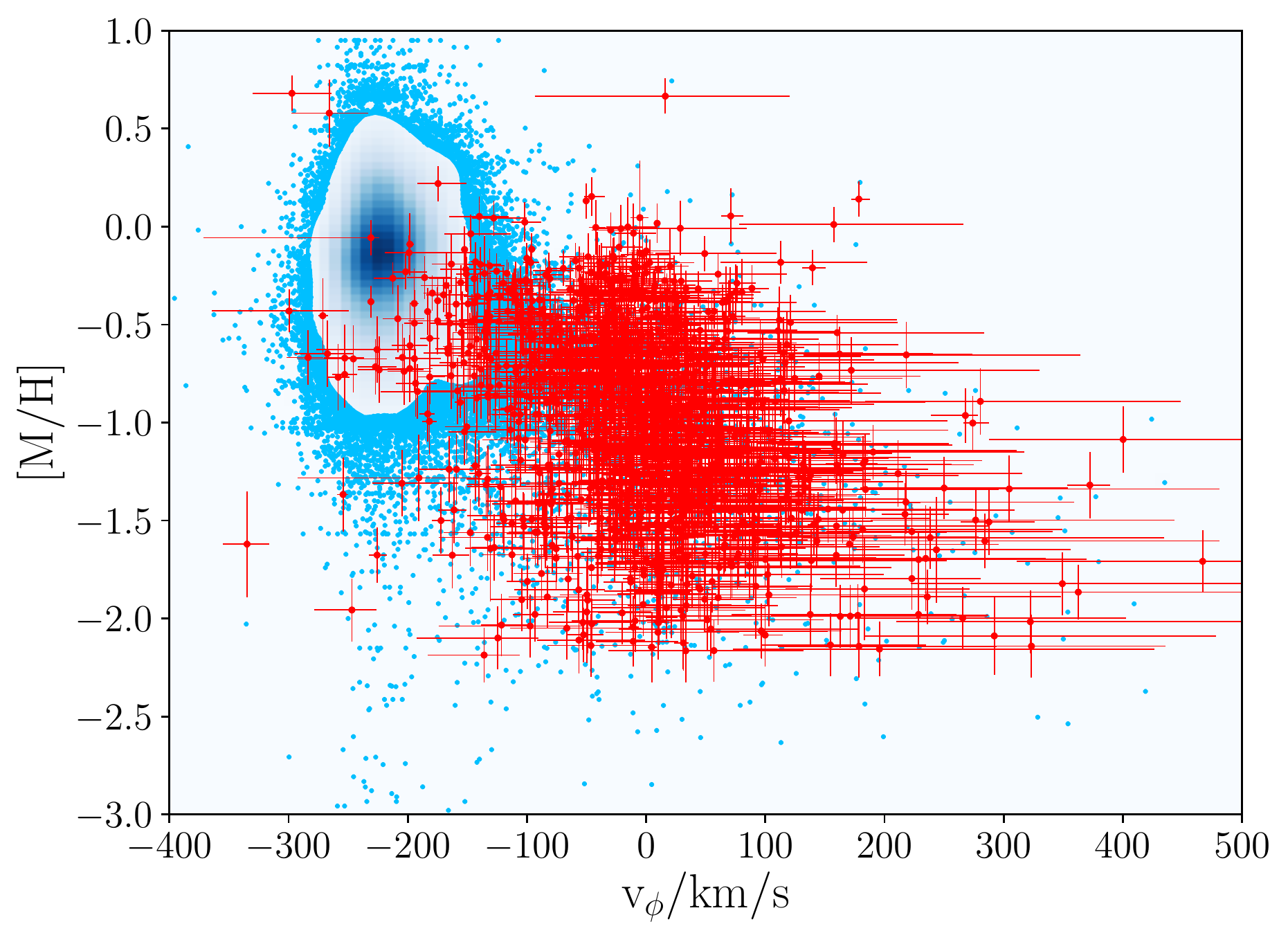}
\caption{Distribution of stars in the dynamically selected stellar
  halo (red) on the metallicity--azimuthal velocity space. For
  comparison, all stars in our TGASxRAVE sample are also shown
  (cyan).}
\label{fig:vphi-met}
\end{figure}

Figure~\ref{fig:vphi-met} shows the distribution of the
dynamically selected halo stars in the $\vphi-[\rm M/H]$ plane: the
stars scatter around $\vphi\sim 0$ and ${\rm [M/H]}\sim -1$~dex, with
a slight but not significant tendency of retrograde motions at lower
metallicities and vice versa \citep[but see also][]{Carollo+2007,Beers+2012,
Kafle+2017}. This plot is, however, severely affected by measurement errors
making trends difficult to discern because of the blurring induced
especially by the uncertainties in azimuthal velocity. Moreover, we
note here that even if we made sure that the error distributions in
azimuthal velocity for our selection of halo stars do not deviate
significantly from Gaussian (the interval of 16th--84th percentiles
is close to two standard deviations), symmetric parallax (and/or
distance) errors typically translate into asymmetric errors in $\vphi$
\citep[e.g.][]{Ryan1992,Schonrich+2014}, which complicates the
understanding of the kinematics of the local stellar halo even
further.

\begin{figure*}
\includegraphics[width=\textwidth]{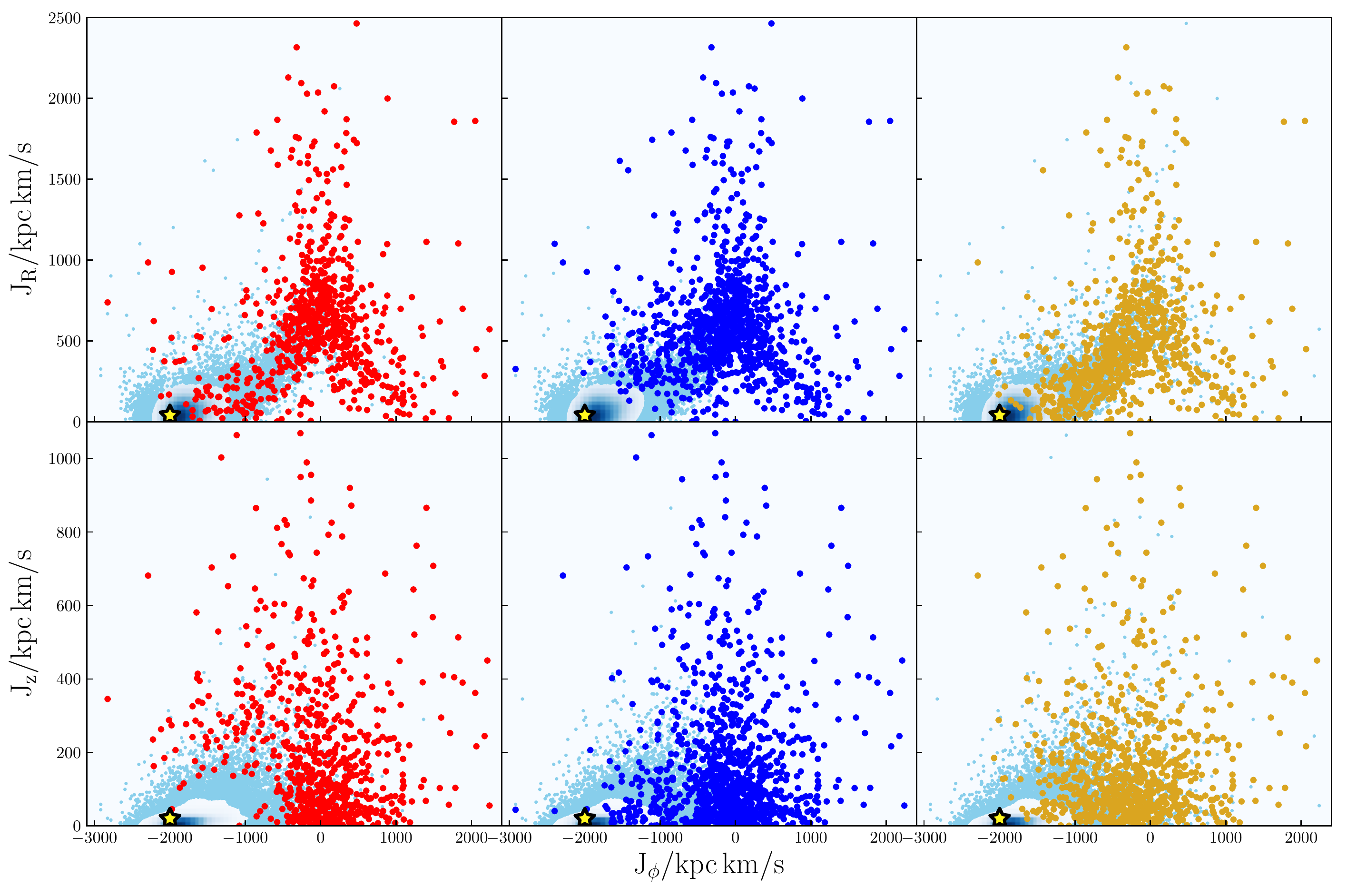}
\caption{Action space distribution of the dynamically selected (red),   kinematically selected (blue),                 and metallicity selected (yellow) stellar halo stars. The yellow
         star marks the location of the Sun.}
\label{fig:act_space}
\end{figure*}

\subsubsection{Distribution in action space}

As shown in Figure~\ref{fig:act_space}, most of the halo stars in action space are
centred around angular momenta $\Jphi\sim 0$, whereas disc stars have high angular
momentum ($\Jphi\sim\Jsun$) and low radial and vertical action ($\Jr\sim\Jz\sim 0$).
It is in this space that the dynamical selection acts, and as discussed before it
depends on the phase space densities of each Galactic component. The ratio
of the phase space densities of the halo $\fhalo$  to that of the discs $\fthin+\fthick$
determines the number of stars assigned to either component in a given volume of
action space. For instance, in a neighbourhood of $(\Jr,\Jphi,\Jz)=(0,\Jsun,0)$ the
ratio of these phase space densities is so low that the probability of assigning a star
to the stellar halo component is in fact negligible.
The probability of halo membership given by Eq.~\eqref{eq:Pcomp} is also low for
$\Jphi<-600$ kpc km/s and $\Jz<200$ kpc km/s, where only a handful of stars satisfy
the criteria. The  likely consequence of this is that some genuine halo stars are
missing from the dynamically selected sample and have been misclassified as thick
disc stars. This is clearly dependent on the dynamical model of the Galaxy assumed
for the selection and it is the reason why the distribution of dynamically selected
halo stars is not symmetric (about $\Jphi=0$) in the $\Jphi-\Jz$ projection, as can also
be seen  from a comparison to the metallicity selected sample.

For the catalogue of stars used in the present study the most
important parameters determining the membership of stars at high
angular momentum $\Jphi$ and low vertical action are the parameters
governing the vertical velocity dispersion of the thick disc:
$\sigma_{z,0}$ and $R_{\sigma,z}$, the vertical velocity dispersion at
the origin and the scale-length of its exponential decay,
respectively. For instance, we find that with a smaller $R_{\sigma,z}$
(which implies a thinner thick disc), the distribution of
dynamically selected halo stars becomes significantly more symmetric in
the $\Jphi-\Jz$ plane. However, \cite{Piffl+2014} show that such a
model provides an overall poorer fit to the observed velocity moments
of all the RAVE stars, and thus it is unlikely to be the solution. We
could in principle search the parameter space for the best model
representing the RAVE data, which also yields a symmetric distribution
of dynamically selected halo stars in action space, but it seems this
exercise should be done when a larger, more complete dataset becomes
available.

For comparison, we also show in Fig.~\ref{fig:act_space} the
distribution in action space for the halo stars selected by kinematics
(middle panels) and metallicity  (right panels). These plots show
that in both cases there are many more halo stars in these two
selections close to $(\Jr,\Jphi,\Jz) = (250, -1000, 100)$ kpc km/s, in
comparison to the dynamical selection.  The stellar halo samples
selected by kinematics and metallicity have, respectively, net angular
momenta $\overline{J_\phi}=-150\pm 13$ and $\overline{\Jphi}=-230\pm
20$ kpc km/s. On the other hand, we find no clear evidence of net
rotation of the dynamically selected stellar halo
($\overline{J_\phi}=-20\pm 15$ kpc km/s). We note, however, that when we try to account for the halo
stars that we might be missing at prograde $\Jphi$ and $\Jz<200$ kpc
km/s (see Fig.~\ref{fig:act_space}), which we do by `mirroring' the
retrograde stars with $\Jz<200$ hence making the $\Jphi-\Jz$ diagram
symmetric at low vertical actions, we get a slightly prograde net
rotation, $\overline{J_\phi}=-35\pm 15$ kpc km/s, which is marginally
compatible with the rotation signal of the inner halo found by other
authors \citep[e.g.][]{Deason+2017}.

Whether the stellar halo is rotating and whether it is made of several
components with different kinematics are still matters of open
debate. However, as we have just shown, it is important to realize
that the conclusions we draw by analysing a sample of halo stars
could (and often do) depend on the selection criteria used to define
such a sample from a parent catalogue of stars in the Galaxy. 

%
%

\subsection{Tilt of the velocity ellipsoid}
\label{sec:vellip}

We now turn  to the characterization of the velocity dispersion tensor, also called
\emph{velocity ellipsoid}, of the local dynamically selected stellar halo, i.e.
\begin{equation}
\label{def:vellip}
\sigma_{ij}^2 \equiv \langle(v_i-\langle v_i \rangle)\,(v_j-\langle v_j \rangle)\rangle,
\end{equation}
where $\langle Q \rangle \equiv \left(\int\de\bv f\,Q \right)/\left(\int\de\bv f\right)$
is the average of the quantity $Q$ weighted by the phase-space DF \cite[e.g.][]{BT08}.
The diagonal elements of this tensor are the velocity dispersions in the three orthogonal
directions $\sigma_{i}^2\equiv\sigma_{ii}^2$, while the off-diagonal elements represent
the velocity covariances.

We employ a three-dimensional multivariate normal distribution $\mathcal{N}(\overline{\bv},
\sigma_{ij})$ to describe the distribution
of velocities of the stars in the stellar halo. This model is characterized by three
mean orthogonal velocities $\overline{\bv}$ and the (symmetric) velocity dispersion tensor
$\sigma_{ij}$, hence nine free parameters. The velocity correlations
\begin{equation}
\label{def:rho}
\rho_{ij}\equiv \frac{\sigma_{ij}^2}{\sigma_i\sigma_j}
\end{equation}
are then related to the angle given by
\begin{equation}
\label{def:alpha}
\tan(2\alpha_{ij})\equiv\frac{2\rho_{ij}\sigma_i\sigma_j}{\sigma_i^2-\sigma_j^2}
,\end{equation}
where $\alpha_{ij}$ is the angle between the $i$-axis and the major axis of the ellipse
formed by projecting the velocity ellipsoid on the $i$-$j$ plane \citep[see][their
Appendix A]{BM98,Smith+2009}. In computing the posterior distributions for the model
parameters, we also take into account measurement errors by
defining the likelihood of a model as the convolution of $\mathcal{N}(\overline{\bv},
\sigma_{ij})$ with the star's error distribution $\gamma(\bv)$. The latter is itself
multivariate normal, hence their convolution results in a multivariate normal with mean the
sum of the means and correlation the sum of the correlations. Given this likelihood, we
finally sample the posterior distribution of the nine parameters with a Markov chain Monte
Carlo method \citep[MCMC; we use the \texttt{python} implementation by][]{Foreman+2013}
assuming uninformative (flat) priors for all the parameters\footnote{
We also tried using log-normal priors for the velocity dispersions since they
are strictly positive, and found the results not to be sensitive to this particular choice.
}.

All our chains easily converge after the so-called \emph{burn-in} phase, and the sampled
posteriors are all limited with no correlation between any parameter couple. Thus, we
derive a maximum likelihood value for all the parameters and estimate their uncertainties
by computing the interval of 16th--84th percentiles of the posterior distributions.
We summarize these results in both a cylindrical and a spherical reference frame in
Table~\ref{tab:mcmc}.

\begin{table}
\caption{Parameters of the local velocity ellipsoid for the dynamically selected halo.}
\label{tab:mcmc}
\renewcommand{\arraystretch}{1.2}
\begin{tabular}{lcclc}
 & Cylindrical & & & Spherical \\
\hline
$\overline{v_R}$ & $-10 \pm 9$ km/s & $\quad$ & $\overline{v_r}$ & $-9 \pm 8$ km/s \\
$\overline{v_z}$ & $-8 \pm 9$ km/s & $\quad$ & $\overline{v_\theta}$ & $7 \pm 8$ km/s \\
$\overline{v_\phi}$ & $-9 \pm 7$ km/s & $\quad$ & $\overline{v_\phi}$ & $-9 \pm 7$ km/s \\
 & & & & \\
$\sigma_R$ & $141 \pm 6$ km/s & $\quad$ & $\sigma_r$ & $142 \pm 6$ km/s \\
$\sigma_z$ & $94 \pm 4$ km/s & $\quad$ & $\sigma_\theta$ & $89 \pm 4$ km/s \\
$\sigma_\phi$ & $78 \pm 4$ km/s & $\quad$ & $\sigma_\phi$ & $74 \pm 6$ km/s \\
 & & & & \\
$\rho_{Rz}$ & $-0.14 \pm 0.06$ & $\quad$ & $\rho_{r\theta}$ & $0.01 \pm 0.06$\\
$\rho_{R\phi}$ & $0.07 \pm 0.06$ & $\quad$ & $\rho_{r\phi}$ & $0.06 \pm 0.06$\\
$\rho_{\phi z}$ & $-0.05 \pm 0.06$ & $\quad$ & $\rho_{\phi\theta}$ & $-0.07 \pm 0.06$ \\
\end{tabular}
\vspace*{15pt}
\end{table}


We  find all the mean velocities to be consistent with zero, and the
velocity dispersions we derive are broadly consistent with previous
measurements \citep[at the high end  of the estimates of
e.g.][]{ChibaBeers2000,Bond+2010,Evans+16}. The velocity ellipsoid is
mildly triaxial ($\sigma_r>\sigma_\theta\gtrsim\sigma_\phi$) and 
the stellar halo is locally radially biased: the \emph{anisotropy
  parameter}
$\beta\equiv1-(\sigma_\theta^2+\sigma_\phi^2)/2\sigma_r^2=0.67 \pm
0.05$.  No significant correlation is found between the velocities in
spherical coordinates, meaning that the halo's velocity ellipsoid is
aligned with a spherical reference frame, which has implications on
mass models of the Galaxy \citep[e.g.][and Sect.~\ref{sec:implic}]
{Smith+2009,Evans+16}. We find a slightly larger vertical velocity
dispersion ($\sigma_z$, $\sigma_\theta$) than previous works;  this
is likely because of the dynamical selection that we have applied,
which is meant to select only stars that are genuinely not moving in
the disc and thus  on orbits with typically large vertical
oscillations. If
we repeat the above analysis for the metallicity selected sample as in
\cite{Helmi+2017}, we have an indication that this is happening, and 
obtain values that are more in line with previous estimates: $(\sigma_r,
\sigma_\theta, \sigma_\phi)=(136 \pm 6, 74 \pm 4,96 \pm 5)$ km/s. 
Also, the large vertical and the small azimuthal dispersion
that we derive using our dynamically selected sample results in a 
velocity ellipsoid that is more elongated on the vertical, rather than
the azimuthal direction \citep[cf.][]{Smith+2009a,Bond+2010,Evans+16}.

\begin{figure}
\includegraphics[width=.5\textwidth]{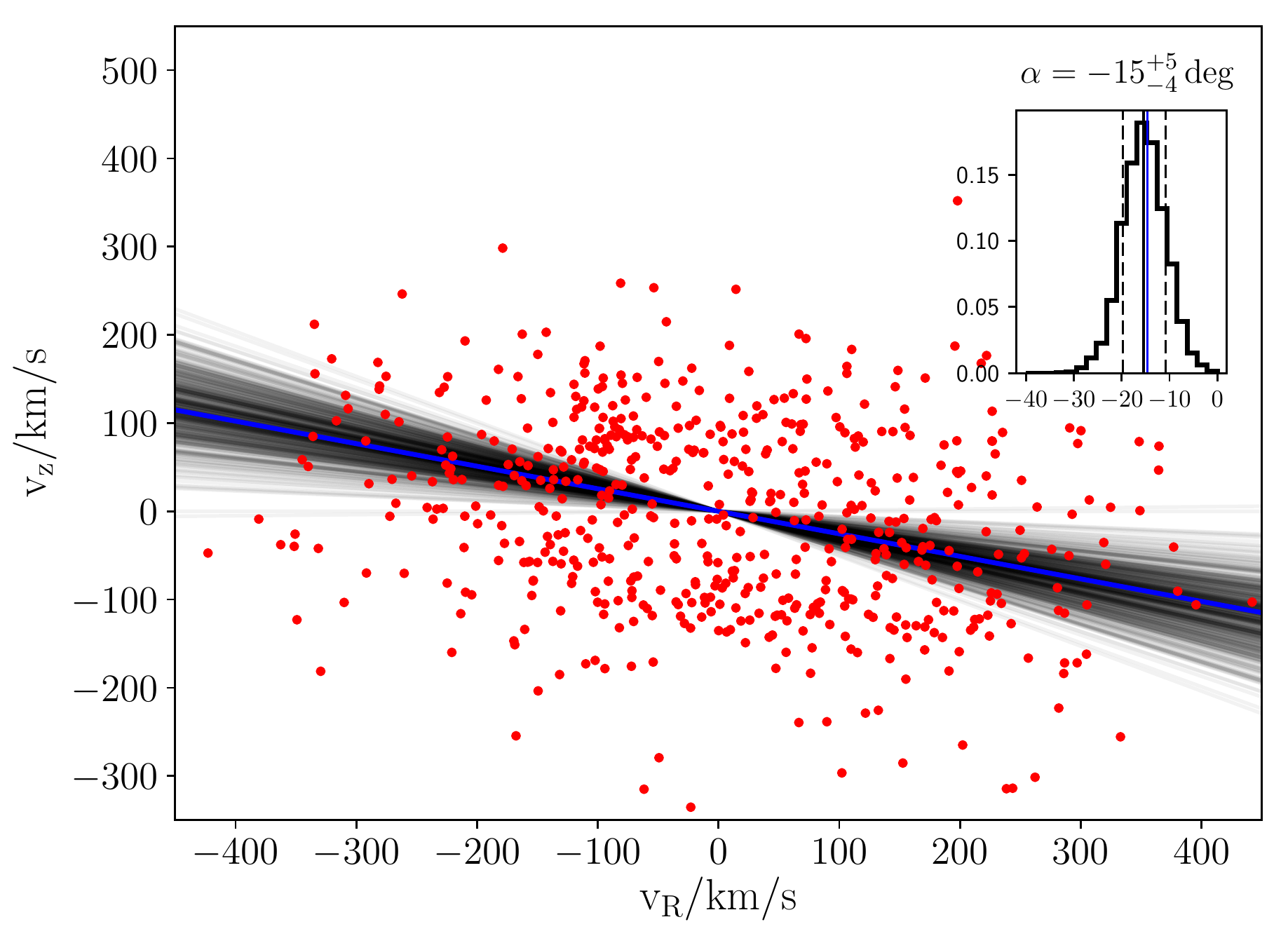}
\caption{Distribution of stars with $-5\lesssim z/{\rm kpc}\leq-1$ in the dynamically selected stellar halo on
                 the $v_R$--$v_z$ subspace. Each black line represents the orientation of the
         velocity ellipsoid in this subspace as given by one sample in the MCMC chain.
         The inset shows the marginalized distribution of the angular slope of the black
         lines (tilt-angle). The blue line marks the expected orientation for a
         spherically aligned velocity ellipsoid.}
\label{fig:tiltangle}
\end{figure}

In Figure~\ref{fig:tiltangle} we show the radial and vertical
cylindrical velocities for the halo stars in our sample located at
$-5\lesssim z/{\rm kpc}\leq-1$ below the disc plane. There is a strong
significant correlation between $\vr$ and $\vz$ with an angular slope
of $\alpha_{Rz}=-15^{+5}_{-4}$ deg, sometimes also called
\emph{tilt-angle}, which is estimated by marginalizing the posterior
probability over all the other model parameters.  It is interesting to
compare this estimate to the value expected for a spherically aligned
velocity ellipsoid, in which case $\alpha_{Rz, \,{\rm
    sph}}=\arctan(z/R)$. For this subset of halo stars we derive
$\alpha_{Rz, \,{\rm sph}}=\arctan(z_{\rm med}/R_{\rm med})\simeq
-14.7$ deg, where $z_{\rm med}\simeq -2$ kpc and $R_{\rm med}\simeq 7.6$ kpc
are their median height and polar radius, respectively. 
The value obtained for $\alpha_{Rz, \,{\rm sph}}$
is therefore consistent with our measurement of the tilt angle $\alpha_{Rz}$.

\begin{figure*}
\includegraphics[width=.49\textwidth]{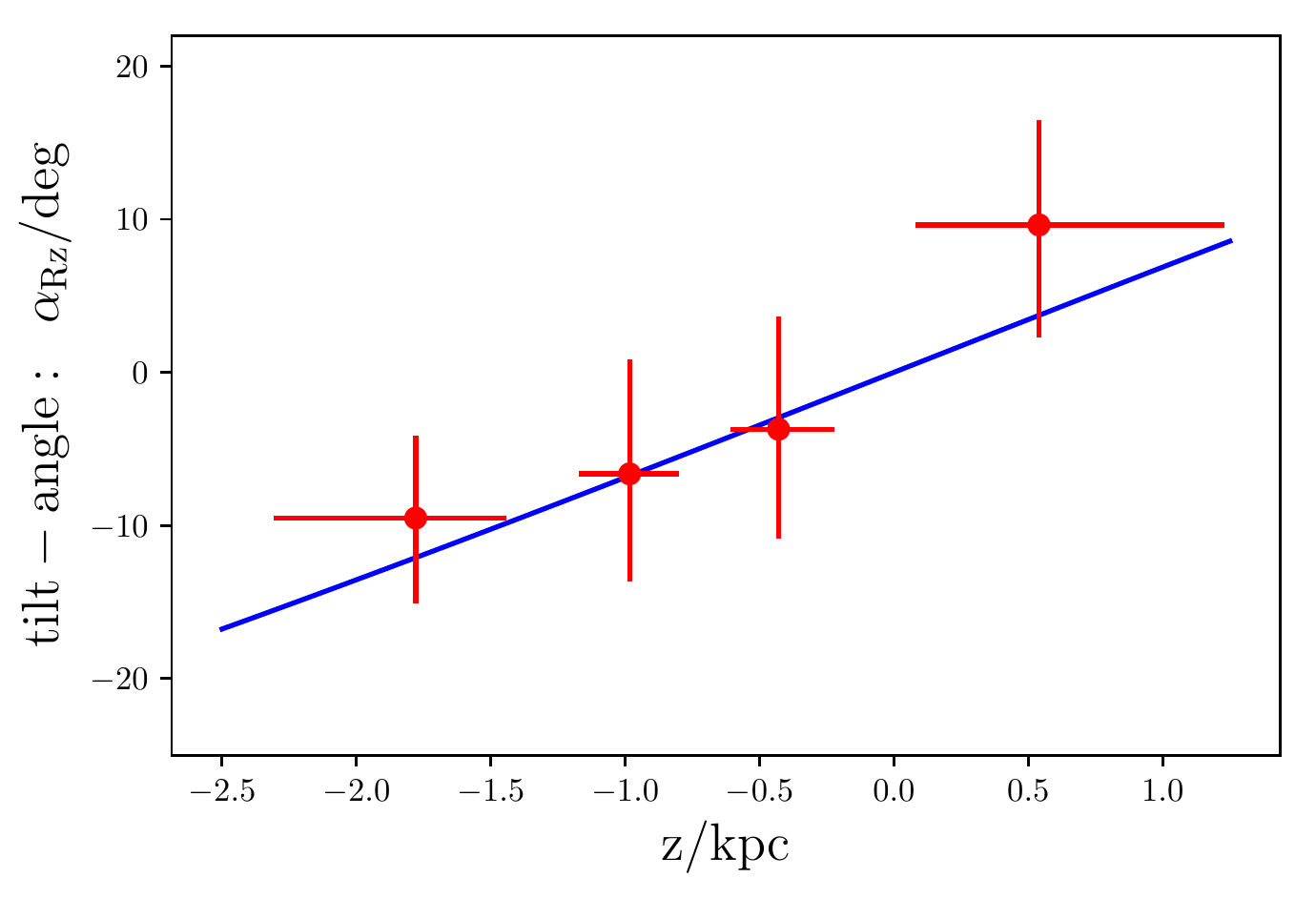}
\includegraphics[width=.49\textwidth]{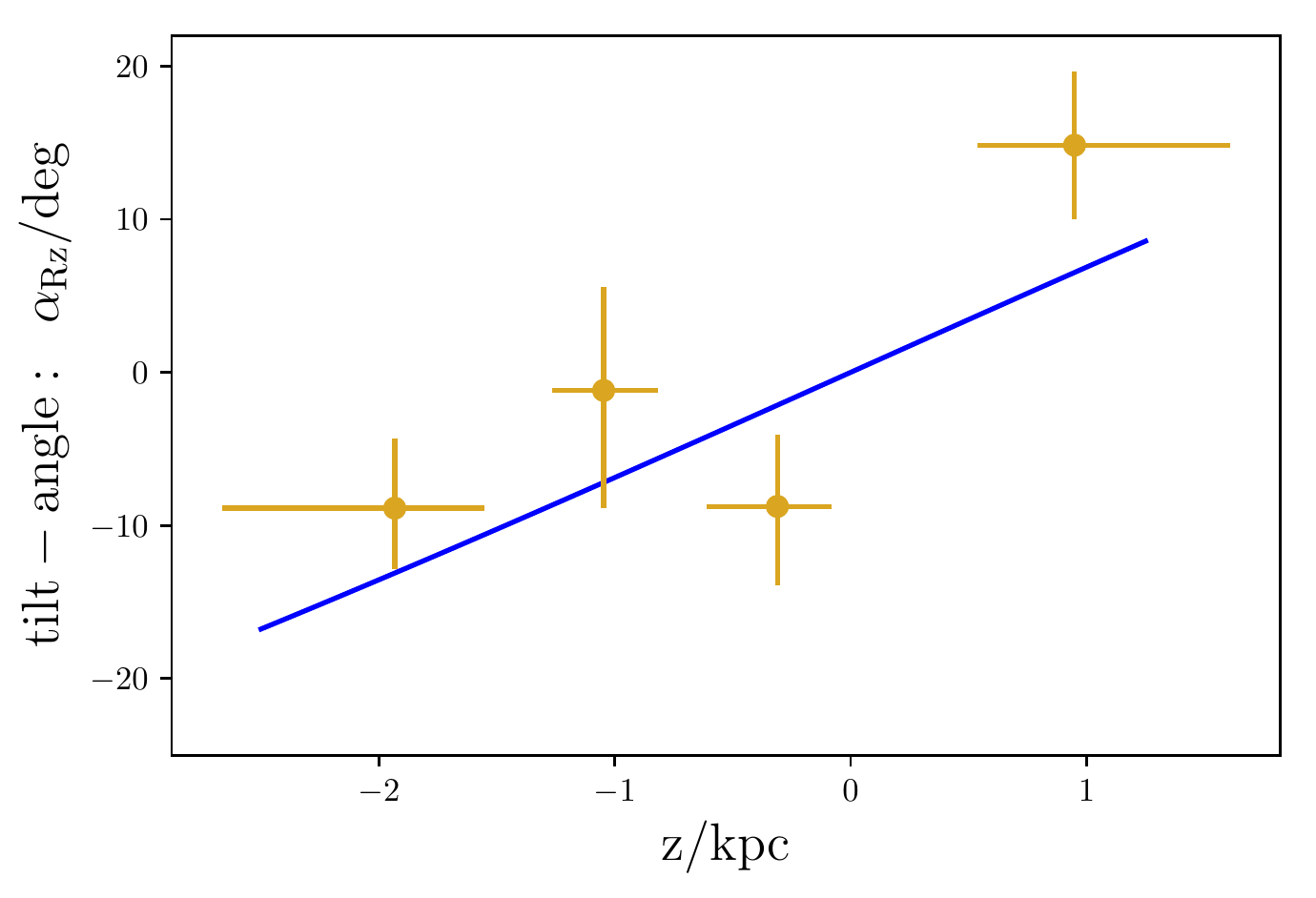}
\caption{Tilt-angle $\alpha_{Rz}$ as a function of height above the Galactic plane for
         the dynamically selected stellar halo (red circles, left) and  the
         metallicity selected stellar halo (yellow circles, right).
         Each bin in $z$ contains roughly the same number of stars ($\simeq 185$ and
         $\simeq 175$ respectively for the two halos).
         The value of $\alpha_{Rz}$ and its uncertainty
         are estimated using a multivariate normal model as in Sect.~\ref{sec:vellip},
         while the error bars in $z$ indicate the 16th--84th percentile of the $z$
         distribution in that bin. The blue solid line indicates a model in which the
         halo's velocity ellipsoid is aligned with the spherical coordinates.}
\label{fig:tiltangle_vs_z}
\end{figure*}

To test whether the halo's velocity ellipsoid is consistent with being
spherically aligned for all $z$, we group the local
($d\leq 3$ kpc) halo stars as a function of their $z$-height in four bins
with roughly 185 stars each, and we use again a multivariate normal
model to determine the tilt-angle $\alpha_{Rz}$.  In
Figure~\ref{fig:tiltangle_vs_z} we plot the most probable
$\alpha_{Rz}$ and its uncertainty, as a function of $z$: there is a
clear close-to-linear correlation in the sense that the tilt-angle is larger 
for samples of stars at greater heights above the Galactic plane. Such
a correlation is precisely what is expected in the case of a 
spherical gravitational potential. An oblate/prolate St\"{a}ckel potential
would result in a shallower/steeper $\alpha_{Rz}-z$ relation 
\citep[see e.g. Eq.~(17) in][]{Budenbender+2015}.

This result is surprisingly robust against different selection criteria
used to identify halo stars. In the right panel of
Fig.~\ref{fig:tiltangle_vs_z} we show, for instance, $\alpha_{Rz}$ as
a function of $z$ for the metallicity selected halo. Furthermore,  in this case
we find that $\alpha_{Rz}$ increases almost linearly with $z$,
consistently with a model in which the velocity ellipsoid is
everywhere spherically aligned. This indicates that  the resulting velocity ellipsoid has a
similar orientation which truly reflects the shape of the underlying
total gravitational potential, even though  selecting stars
by their orbits or by metallicity yields somewhat
different local velocity distributions for the halo (see
Sect.~\ref{sec:localkin}). These results are also robust to
different distance cuts. 

\subsubsection{Implications for mass models of the Galaxy}
\label{sec:implic}

The measurement of the orientation of the velocity ellipsoid of the
stellar halo yields crucial insights on the overall gravitational
potential of the Galaxy. While its axis ratios are related to the
halo's anisotropy and hence possibly to its formation history, at each
point in space the orientation of the halo's velocity ellipsoid is
directly related to the shape of the total gravitational potential. For example, it has long been known that if the potential is
separable in a given coordinate system, then its velocity ellipsoid is
oriented along that coordinate system
\citep[e.g.][]{Eddington1915,LyndenBell62}, but recently it has been
shown that  the contrary also holds, i.e. if the velocity ellipsoid is
everywhere aligned with a given coordinate system, then the potential
is separable in those coordinates \citep{Evans+16}. The necessary and
sufficient condition for the potential of a steady-state stellar
system to be separable in some given St\"{a}ckel coordinates
$(q_1,q_2,q_3)$, i.e. it is a St\"{a}ckel potential, can be rewritten
using the Jeans equations as follows: \emph{If} i) \emph{all the
  velocity dispersions in a given orthogonal frame are different}; ii)
\emph{all the second-order mixed moments of the velocity vanish,
  $\langle v_i v_j \rangle=0$ for $i\neq j$; and} iii) \emph{all
  fourth-order mixed moments with odd powers vanish, $\langle v_1^l
  v_2^m v_3^n \rangle=0$, where $l+m+n=4$ and at least one of $l,m,n$
  is odd} \citep{AnEvans2016}.  For our dynamically selected halo
sample, the correlation coefficients $\rho_{ij}$ with $i\neq j$
(Eq.~\ref{def:rho}), which are the second-order mixed velocity moments
normalized to the respective velocity dispersions, are indeed
compatible with being null in spherical coordinates.  We also compute
the fourth-order moments numerically\footnote{
\begin{equation*}
\langle v_i v_j v_p v_q \rangle = \frac{1}{N_{\rm stars}}\sum_{n=0}^{N_{\rm stars}}
        v_{n,i} v_{n,j} v_{n,p} v_{n,q}
\end{equation*}
for $i,j,p,q$ being any spherical coordinate and $v_{n,i}$ the $i$-th
velocity component of the $n$-th star.  } and we derive fourth-order
correlation coefficients by normalizing them by the respective
velocity dispersions (squared), as in Eq.~\eqref{def:rho}. We
compute their uncertainties using the Monte Carlo samples.  We find
values no higher than $0.05 \pm 0.06$ for the normalized mixed moments
with odd powers in velocity and of the order of $0.33 \pm 0.06$ for
the normalized mixed moments of the type $\langle v_i^2 v_j^2
\rangle$.

These results indicate that at least \emph{locally} the stellar halo
of the Galaxy  respects the conditions of the theorem by
\cite{AnEvans2016}, hence suggesting that the total Galactic potential
may be separable in spherical coordinates.  This is intriguing, since
the only axisymmetric potentials which yield a finite mass for the
Galaxy and are separable in spherical coordinates are the spherical
potentials \citep[e.g.][]{Smith+2009}. We note, however, that it is
also possible to construct physically plausible Galaxy models in which
the velocity ellipsoid is locally aligned with the spherical
coordinates even in an oblate \citep{BinneyMcMillan2011} or a triaxial
gravitational potential \citep{Evans+16}, hence no strong conclusions
can be reached with a {\it local} halo sample like the one we are
using here.  On the other hand, in non-spherical potentials the
misalignment of the velocity ellipsoid with the spherical coordinates
is typically small only in a limited region of space and, intuitively,
 gets smaller when the  potential is closer to spherical symmetry
\citep{Evans+16}. Future \Gaia~data releases will probe unexplored
regions of the Galactic stellar halo and will lead to a large increase
in the size of samples of halo stars, therefore allowing us to put significant
constraints on the geometry of the Galactic potential.

\section{Conclusions}
\label{sec:concl}

In this paper we have devised a novel method to select stars in the
stellar halo of our Galaxy from a catalogue with measured positions
and velocities. Our method involves characterizing the orbit of each
star using its actions and then computing the probability that the star
is a member of a given Galactic component employing a specified
self-consistent dynamical model for the Galaxy. We have applied this
method to the $\sim 175$k stars found in the intersection of the TGAS
and RAVE catalogues and we have used the DF-based dynamical model
by \cite{Piffl+2014}.  We have compared this \emph{dynamically based}
selection of halo stars with one based purely on kinematics and one
based on  metallicities. We summarize here the main findings of
the paper:
\begin{itemize}
\item[(i)] We find 1156 halo stars in the solar neighbourhood, a number  comparable  to those found using the 
   two alternative selection methods. 
\item[(ii)] The halo stars in our sample are typically
        distant giants on very elongated orbits; they are mostly more metal-poor than
    [M/H]$\lesssim -0.5$, but we find that roughly half of the sample has  
    [M/H]$\ge -1$, in broad agreement with what is found for a kinematically selected sample. 
  \item[(iii)] The velocity distribution of the dynamically selected
    sample is reasonably Gaussian,  with means
    $(\overline{v_r}, \overline{v_\theta}, \overline{v_\phi}) = (-9
    \pm 8, 7 \pm 8, -9 \pm 7)$ km/s and dispersions of
    $(\sigma_r, \sigma_\theta, \sigma_\phi) = (142 \pm 6, 89 \pm 4, 74
    \pm 6)$ km/s. A kinematically selected sample is affected by strong
    biases and is clearly non-Gaussian, while a sample based on metallicity is also well-fit by a
    multivariate normal distribution, but shows slightly positive
    prograde rotational motion; the velocity ellipsoid has a
    slightly different shape, with $(\sigma_r,\sigma_\theta,
    \sigma_\phi)=(136 \pm 6, 74 \pm 4,96 \pm 5)$~km/s.
  \item[(iv)] Differences in the properties of the velocity
    distributions can be traced back to the criteria used by the
    different selection methods. A dynamically based method suppresses
 the contribution of stars that have close to thick disc-like orbits,
    while if based on metallicity these stars are selected provided
    they have [M/H]$<-1$ dex (which leads to a higher $\sigma_\phi$ and
    a lower $\sigma_\theta$).
  \item[(v)] Despite these differences, for both the dynamical and the metallicity-based  selection methods,
    the tilt angle of the stellar halo is robustly measured and is consistent with
    being spherically aligned. 
\item[(v)]   All the second-order velocity mixed moments, together
        with all the fourth-order ones with odd velocity powers, are consistent with being
    null in volume probed by the catalogue, which suggests that the total gravitational potential of the Galaxy
    has to be locally close to spherical \citep{AnEvans2016}.  
\end{itemize}

The method described here can be easily extended to categorize a
sample of stars for which at least one  of the six-dimensional
phase-space coordinates is missing. Given a dynamical model in the
form of a distribution function $f(\bJ)$ we can always define
membership probabilities as in Sect.~\ref{sec:membership} by replacing
the value of the six-dimensional phase-space density $f_\eta(\bJ)$
for each Galactic component $\eta$ by its analogue obtained by
marginalizing over the missing coordinate. In the near future
\Gaia~will provide five-dimensional information for more than a
billion stars in the Galaxy, but radial velocities will be available
for just about $10\%$ of the total, hence devising algorithms that can
exploit the full extent of the information available already with the
second \Gaia~Data Release is of vital importance.

\begin{acknowledgements}
We acknowledge  financial  support from a VICI grant from the Netherlands
Organisation for Scientific Research (NWO).
This work  has  made  use  of  data  from  the  European  Space  Agency  (ESA)  
mission Gaia (\url{http://www.cosmos.esa.int/gaia}), processed by the Gaia Data
Processing and Analysis Consortium (DPAC,
\url{http://www.cosmos.esa.int/web/gaia/dpac/consortium}).
Funding for the DPAC has been provided by national institutions, in particular
the institutions participating in the Gaia Multilateral Agreement.
\end{acknowledgements}


\bibliographystyle{aa} 
\bibliography{ref_dynHalo} 

\appendix
\section{Results with spectro-photometric distances from RAVE DR5}
\label{sec:app}

Here we summarize the results of the selection method described in this paper with regard
to the catalogue of stars where parallaxes are either trigonometric from TGAS or
spectro-photometric from the RAVE DR5 database \citep{Kunder+2017}, determined
with the method by \cite{Binney+2014}. In this case  the stellar parameters,
such as surface gravity and calibrated metallicity, are  obtained from the DR5
pipeline, and not from  the updated \cite{McMillan+2017} estimate. 

\begin{itemize}
\item We get a final sample of 159,702 stars by applying the cuts described in
          Sect.~\ref{sec:data}.
\item The sample of dynamically selected halo stars (see Sect.
      \ref{sec:membership}) is composed of 743 stars, i.e. $0.46\%$ of the total
      sample.
\item 278 halo stars have [M/H]$>-1$ ($37\%$),  128 of which are on retrograde
      orbits ($46\%$); 105 halo stars ($14\%$) are found at relatively low
      velocities $|{\bf V}-{\bf V}_{\rm LSR}|<{\bf V}_{\rm LSR}$.
\item The mean velocities of the local ($d<3$ kpc) dynamically selected stellar halo
      (estimated as in Sect.~\ref{sec:vellip}) are all consistent with being null.
      The velocity dispersions are $(\sigma_r, \sigma_\theta, \sigma_\phi)=
      (149\pm 6, 96\pm 4, 87\pm 5)$ km/s and the velocity correlation coefficients
      are $(\rho_{r\theta}, \rho_{r\phi}, \rho_{\phi\theta})=(0.05\pm 0.06,
      -0.03\pm 0.06, 0.04\pm 0.06)$.
\item For the sample of halo stars at $z\leq -1$ kpc below the Galactic plane, we
      find a tilt-angle $\alpha_{Rz}=-15\pm 5$ deg, consistent with the expectation
      for a spherically aligned velocity ellipsoid of $\alpha_{Rz}=\arctan(
      z_{\rm med}/R_{\rm med})=-16$ deg.

\end{itemize}

\end{document}